\documentclass[final, 5p, times, twocolumn]{elsarticle}

\usepackage{gensymb}
\usepackage{graphicx}
\usepackage[utf8]{inputenc}
\usepackage{url}
\usepackage{bm}
\usepackage{booktabs}
\usepackage{lineno}
\usepackage{ulem}

\journal{Astroparticle Physics}
\begin{document}
\begin{frontmatter}

\title{Particle Identification In Camera Image Sensors Using Computer Vision}

\author[a,b]{Miles Winter\corref{correspondingauthor}} 
\author[a,b]{James Bourbeau\corref{correspondingauthor}}
\author[a,b]{Silvia Bravo} 
\author[b,c]{Felipe Campos} 
\author[a,b]{Matthew Meehan\corref{correspondingauthor}}
\author[e]{Jeffrey Peacock}
\author[a]{Tyler Ruggles} 
\author[a,b]{Cassidy Schneider} 
\author[d]{Ariel Levi Simons}
\author[a,b]{Justin Vandenbroucke}

\cortext[correspondingauthor]{Corresponding authors: \url{winter6@wisc.edu}, \url{jbourbeau@wisc.edu}, \url{mrmeehan@wisc.edu}}

\address[a]{Department of Physics, University of Wisconsin-Madison, Madison, WI 53706, USA}
\address[b]{Wisconsin IceCube Particle Astrophysics Center, Madison, WI, 53703, USA}
\address[c]{University of California, Berkeley, Berkeley, CA, 94720, USA}
\address[d]{University of Southern California, Los Angeles, CA, 90007, USA}
\address[e]{Sensorcast, Boulder, CO 80305, USA}

\begin{abstract}
We present a deep learning, computer vision algorithm constructed for the purposes of identifying and classifying charged particles in camera image sensors. We apply our algorithm to data collected by the Distributed Electronic Cosmic-ray Observatory (DECO), a global network of smartphones that monitors camera image sensors for the signatures of cosmic rays and other energetic particles, such as those produced by radioactive decays. The algorithm, whose core component is a convolutional neural network, achieves classification performance comparable to human quality across four distinct DECO event topologies. We apply our model to the entire DECO data set and determine a selection that achieves $\ge90\%$ purity for all event types. In particular, we estimate a purity of $95\%$ when applied to cosmic-ray muons. The automated classification is run on the public DECO data set in real time in order to provide classified particle interaction images to users of the app and other interested members of the public.
\end{abstract}


\end{frontmatter}

\section{Introduction}\label{sec_intro}
The ubiquity of smartphone devices worldwide has sparked an explosion in the field of distributed sensors; their widespread adoption has effectively instrumented global population centers with a variety of detectors. The CMOS image sensors in modern smartphones are based on similar semiconductor technology to that found in professional telescopes and particle physics detectors, enabling them to detect cosmic rays and other ionizing charged particles. These particles have long been a background nuisance for CCDs used in astronomical cameras~\cite{groom2002}, however several recent projects including the Distributed Electronic Cosmic-ray Observatory~\cite{vandenbroucke2015} seek to use this background as signal for both scientific and educational purposes. It may be possible for such networks of smartphones to detect extensive air showers created by ultra-high energy cosmic rays (UHECR) above $10^{20}$ eV, if challenging user density targets are met~\cite{crayfis_2016}. This is a powerful and cost-effective way to extend UHECR measurements to higher energies, but there are substantial hurdles to achieving this goal~\cite{unger_2015}. Since it is also possible to detect local radioactivity with camera sensors~\cite{cogliati_2014}, networks of smartphones could be used as radiation monitors. More exotic analyses have also been proposed, such as searching for correlated extensive air showers created when an ultra-high-energy photon interacts with the heliosphere~\cite{credo_2017}. One major hurdle limiting these scientific pursuits is accurate and efficient particle identification, which is necessary to reject the radioactive background for cosmic-ray measurements or vice-versa for radiation measurements. In this paper we describe a computer vision algorithm developed to identify the charged particles detected by camera image sensors. We then apply it to the data set produced by the Distributed Electronic Cosmic-ray Observatory (DECO) \cite{vandenbroucke2015,vandenbroucke2016}, the first publicly available cosmic-ray smartphone application.

\begin{figure}[h]
  \centering
  \includegraphics[width=1\linewidth]{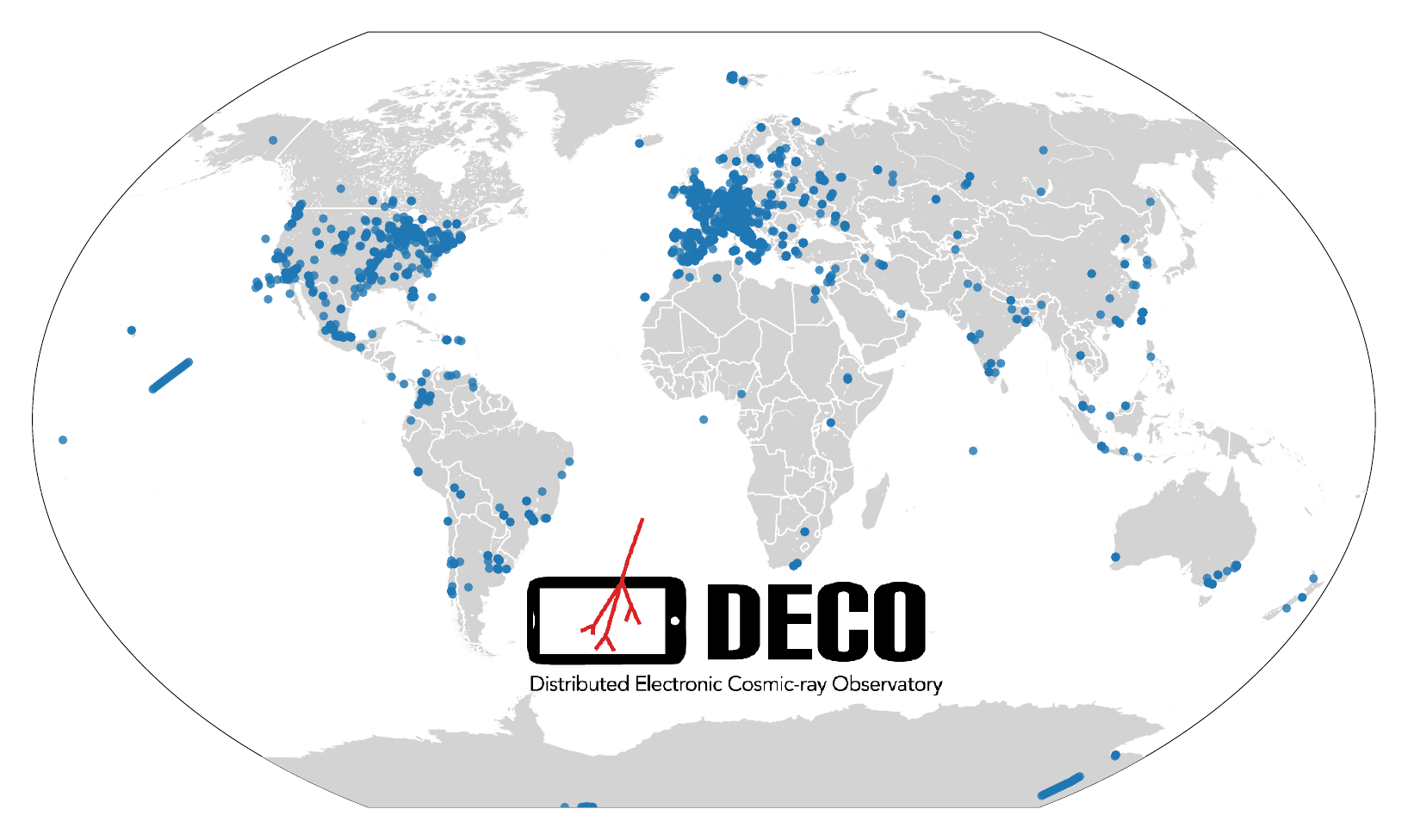}
  \caption{\small{World map showing the global network of DECO users. Dots indicate data taking locations and span 80 different countries. Every continent including Antarctica is represented. Lines of data points, such as those in Antarctica and west of the Americas, indicate users running DECO on plane flights. Map plotted with a Kavrayskiy VII projection and up to date as of December 2017.}}
  \label{map}
\end{figure}

	DECO detects cosmic rays by way of an Android application that began beta testing in October 2012 and was released publicly in September 2014. DECO is designed to detect ionizing radiation that traverses silicon image sensors in smartphones. The resulting dataset consists of images recorded by users worldwide (Figure~\ref{map}) that contain evidence of charged particle interactions. Due to the diverse ecosystem of Android phones on the market, the systematic variation in data taking conditions, and the variety of particle event morphologies, classification of DECO events presents a unique challenge. Our initial work using straight cuts to classify events in the highly heterogeneous dataset was moderately successful in classifying some event types, but identifying a cosmic-ray muon sample with high purity proved challenging. We present a computer vision algorithm based on a convolutional neural network for classifying DECO events. Additional cosmic-ray cell phones apps mentioned above could also benefit from the approach described here. We presented initial results from our CNN classification in~\cite{meehan2017}. More recently, during preparation of this paper, ~\cite{crayfis_cnn} appeared and describes a CNN algorithm intended for use as an online cosmic-ray muon trigger.

\section{DECO App}\label{sec_deco_app}
The DECO detection technique uses similar ionization-detecting semiconductor technology to that found in the silicon trackers of professional particle physics experiments~\cite{ackermann2012,cms2008}. Ionizing charged particles that travel through the sensitive region (i.e. depleted region) of a phone's image sensor are detected via the electron-hole pairs they create. The DECO app, which can be run on any Android device with Android version $\ge$ 2.1, is designed to be run with the camera face down or covered in order to minimize contamination from background light. While running, the app repeatedly takes long-duration ($\sim$50 ms) exposures and runs them through a two-stage filter to search for potentially interesting events. This filter first searches a low-resolution image for $N$ pixels above an intensity threshold, and if passed, analyzes a high-resolution image in the same manner. The intensity is the sum of the red, blue, and green color values (RGB) for each pixel. Images that pass both filters are tagged as ``events'' and are automatically uploaded to a central database for offline analysis. Additionally, the app has a ``minimum bias'' data stream that saves one image every five minutes per device for offline calibration and noise studies. In particular, they are used to determine the appropriate value of $N$ for the online filter to select potentially interesting events. The app’s online filter is simple and efficient in order to maximize livetime, while more detailed analyses of images are performed offline. The DECO data can be browsed using a public website~\cite{deco_data}, where users can perform queries using various metadata including
time stamp (UTC), latitude and longitude (rounded to nearest 0.01$^{\circ}$ for privacy), event vs. minimum bias categorization, Android phone model, and device ID.

\begin{figure}[h]
  \centering
  \includegraphics[width=1\linewidth]{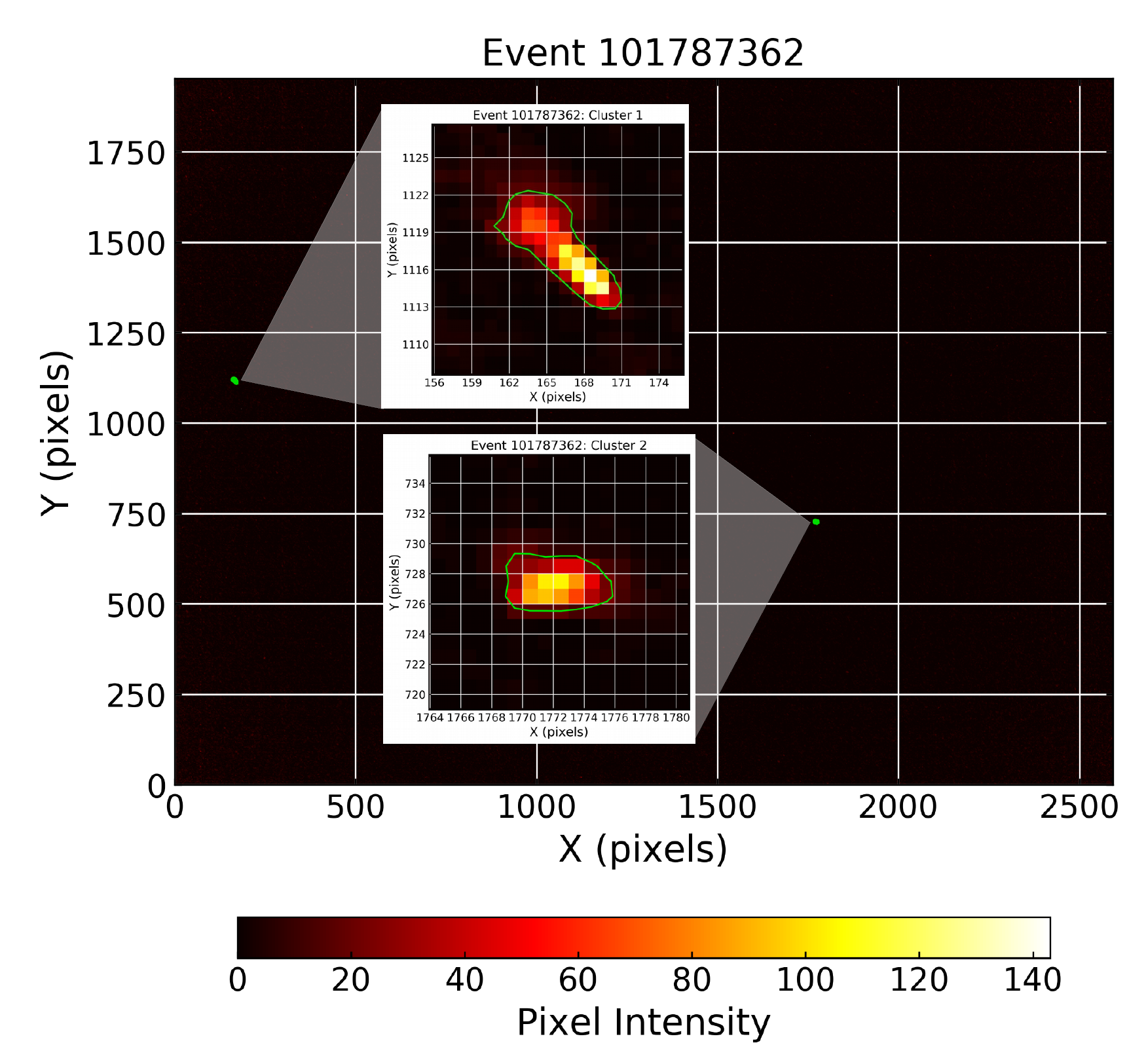}
  \caption{\small{Example of a full camera image that passed online filtering. During offline analysis, a contour-finding algorithm is used to identify hit clusters of pixels. In this event, two clusters (shown with green contours) were identified for further analysis and classification. The color scale represents the pixel intensity, scaled to the brightest pixel, after a conversion to grayscale.}}
  \label{fig_contour}
\end{figure}

Offline analysis of images that pass the app's online filter begins with a contour-finding algorithm to locate clusters of bright pixels. We use the marching squares algorithm, a special case of the marching cubes algorithm \cite{lorensen1987,scikit-image}, to search for groups of at least 10 pixels with a minimum RGB sum of 20. These clusters of pixels are then grouped together at a higher level: any clusters within 40 pixels of one another are considered a single group. This grouping is to account for electrons which can scatter in and out of the camera sensor, creating multiple nearby clusters of pixels with distinct contours. Figure~\ref{fig_contour} shows an example of the contours found in a DECO image with this algorithm. 

\subsection{Event Types}\label{sec_events}
\begin{figure}[h]
  \centering
  \includegraphics[width=1\linewidth]{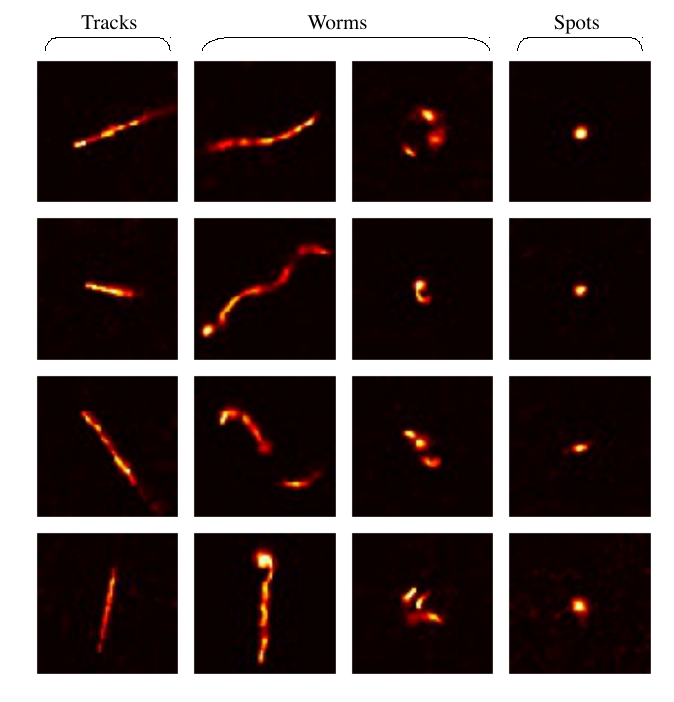}
  \caption{\small{Representative sample of the three distinct types of charged particle events that require classification. Tracks and spots, left and right columns, respectively, are generally observed to have consistent and predictable features. Worms, middle two columns, are observed to have a much wider variety of features, many of which present potential classification confusion when compared to track-like and spot-like features. Each image above has been converted to grayscale and cropped to $64\times64$ pixels.}}
  \label{blob_groups}
\end{figure}

\begin{figure*}[t]
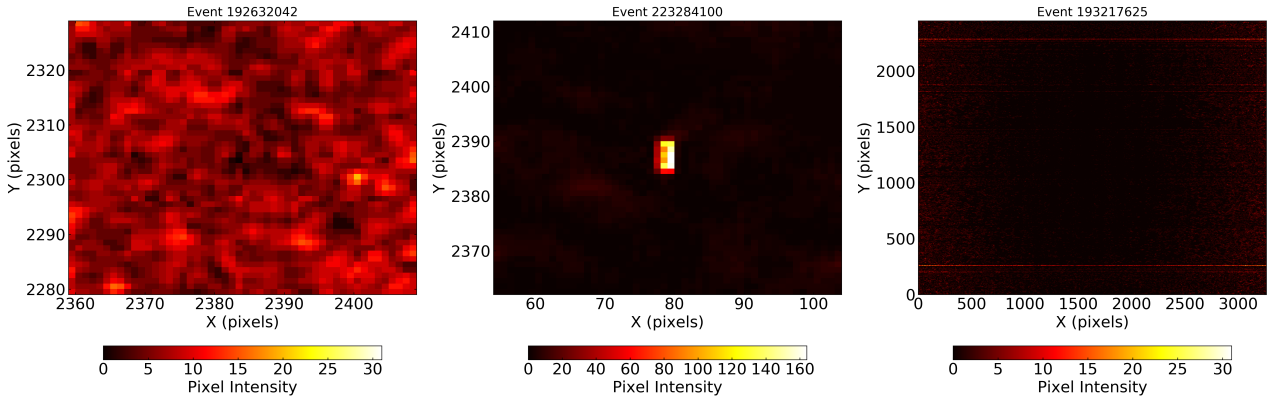

  \centering
  \includegraphics[width=0.3\linewidth]{./plots/noise_zoom.pdf}
  \includegraphics[width=0.3\linewidth]{./plots/hotspot_zoom.pdf}
  \includegraphics[width=0.3\linewidth]{./plots/artifact_full.pdf}
  \caption{\small{Examples of non-particle (noise) events in the DECO dataset. Left: noise due to thermal fluctuations. Center: hot pixels, i.e., pixels that have regular, geometric shapes and typically repeat in the same location. Right: row of bright pixels, likely an artifact of the image sensor readout. The color scale represents the pixel intensity, scaled to the brightest pixel in each image, after a conversion to grayscale.}}
  \label{non_blob_groups}
\end{figure*}

There are three categories of charged particle events in the DECO dataset: tracks, worms, and spots. These are named according to the convention in~\cite{groom2002}, which categorizes events based on their morphology. Tracks are long, straight clusters of pixels in an image created by high-energy (GeV) minimum-ionizing cosmic rays. These are predominantly cosmic-ray muons at sea level and primary cosmic rays (mostly protons) above $\sim$20,000~ft altitude~\cite{PDG}. Worms are named for the curved clusters of pixels caused by the meandering paths of electrons that have undergone multiple Coulomb scattering interactions. These electrons are likely the result of local radioactivity. Worms can also be seen as two or more nearby, disconnected clusters of pixels, which are the result of an electron scattering in and out of the sensitive region of the camera sensor. Spots are smaller, approximately circular clusters of pixels that can be created by various interactions. They are likely predominantly caused by gamma rays that Compton scatter to produce a low energy electron that is quickly absorbed. Spots can also be produced by alpha particles, which also have a very short range in silicon, or by cosmic rays incident normal to the sensor plane. Figure~\ref{blob_groups} shows the characteristic camera sensor response for each of the three interaction signatures detected by DECO. In addition to the three particle interaction categories, there are also events due to light in the sensor occurring when it is not sufficiently shielded, and several categories of noise: hot spots, thermal noise fluctuations, and large-scale sensor artifacts such as rows of bright pixels \cite{vandenbroucke2015}. While non-particle events, shown in Figure~\ref{non_blob_groups}, are not particularly of interest from an analysis standpoint, they do cause potential classification confusion. It is worth noting that these event categories are motivated both by their morphologies and the potential physics analyses that would utilize different categories of events as signal or background. For example, efficient track identification (and worm rejection) is required to detect UHECRs using networks of smartphones or to perform cosmic-ray experiments in a classroom setting. Worms, on the other hand, would need to be identified in order to use DECO or a similar app as a radiation monitoring system.


\subsection{Initial Classification Approach}\label{sec_init_class}
Given the numerous event types, both particle and non-particle, and the increasing number of images being collected by DECO, there is a growing need for a reliable computerized event classification system. However, there are several challenges associated with characterizing the DECO dataset in a way that requires little human intervention. Due to the inhomogeneity in hardware\footnote{Users have run DECO on 604 distinct phone models to date.} and data acquisition conditions, otherwise identical events may be detected differently, for example due to fluctuations in brightness, background noise, or number of pixels hit. Additionally, DECO particle events possess rotational and translational symmetry, which must be accounted for by classification algorithms.

An initial algorithm that classified DECO events used straight cuts applied to geometric metrics that were combined to make a binary classification: track or non-track. Clusters of pixels were identified using the marching squares algorithm described in Section~\ref{sec_deco_app}. The binary classification identified low-noise images with a single cluster of pixels, not containing any sub-clusters (i.e. evidence of an electron scattering out of the sensor plane), with a minimum area of 10 pixels, and an eccentricity $>$0.99, where eccentricity is calculated using image moments as described in ~\cite{image_moments}. The last two requirements were intended to select larger, line-like events, such as tracks. This method accurately distinguished tracks from spots, but struggled to separate tracks and worms, presumably due to their similar morphology. Many worms only curve slightly and have a high eccentricity. These events are unlikely to be high-energy muons due to their curvature, but the classification based on straight cuts could not distinguish them from tracks. Fortunately, advances in the quickly developing field of machine learning offer techniques to overcome these classification challenges. 

\section{Deep Learning}\label{sec_deep_learn} 
\subsection{Background}\label{sec_background}
Deep learning is a subset of machine learning focused on building models that are capable of learning how to describe data at multiple levels of abstraction. This is achieved with a nested hierarchy of simple algorithms that when combined can form highly complex and diverse representations. At each layer of the nested hierarchy, a non-linear transformation of the previous layer's output is typically performed, which results in the deeper layers of the model seeing a progressively more abstract representation of the original input. By learning features at multiple levels of abstraction, the model has the ability to learn complex mappings between the input and output directly from data \cite{bengio2009}. This is particularly advantageous when dealing with higher-level abstractions that humans may not know how to explicitly describe in terms of the available input.

Deep learning models are typically constructed with four basic components in mind: (1) a specific dataset, (2) an objective function\footnote{In the case of minimization, the objective function is commonly referred to as the cost, loss, or error function.}, i.e. the function that will be maximized or minimized, (3) the optimization procedure to be used on the objective function throughout the learning process, and (4) an appropriate structure for the model given the analysis goals and dataset characteristics. For our purposes, a particularly relevant and widely used example of such a model is the feedforward neural network, also known as the multilayer perceptron \cite{rosenblatt62, reed1998}, which can be used to perform a number of tasks, including classification. 

For classification, we begin by assuming that there exists some function, $f^*$, that describes the true mapping between input vector, $\bm{x}$, and category, $y$, such that $y=f^*(\bm{x})$. In this case, the goal of the feedforward neural network is to construct a mapping, $\bm{y}=f(\bm{x};\bm{\theta})$, then learn which value of the parameter vector, $\bm{\theta}$, provides the best approximation between $f^*$ and $f$~\cite{goodfellow2016}. The categorical label, $\bm{y}$, is a unit vector containing all zeros, except for the index that corresponds to the $y$th category in the model, which has a value of 1. The function $f$ is typically a series of nested functions, $f(\bm{x})=f_n(f_{n-1}(...f_2(f_1(\bm{x}))...))$, with depth $n$, where $f_1$ corresponds to the input layer, $f_2$ through $f_{n-1}$ are hidden layers\footnote{Intermediate layer outputs are always connected as inputs for other layers and are therefore never visible as network outputs, hence the term ``hidden''.}, and $f_n$ is the layer that provides the desired output (e.g. probabilities for input $\bm{x}$ belonging to each individual category in $\bm{y}$). 

Each layer consists of a specified number of units, called neurons, that each compute a weighted linear combination of the inputs followed by a non-linear function which outputs a single, real-valued input for the next layer. Traditionally, layers have a dense, fully connected structure where the output of each neuron in a given layer is connected as input to all the neurons in the next layer. In this case, the output of the $n$th layer, $\bm{x}_n$, has the following vector representation:
\begin{equation}
\bm{x}_n = g(\bm{W}_n\bm{x}_{n-1}+\bm{b}_n),\label{neuron}
\end{equation}
where $\bm{x}_{n-1}$ is the output of the previous layer's neurons, $\bm{W}_n$ is a matrix of weights, $\bm{b}_n$ is a vector of biases, and $g$ is the non-linear function, also known as the activation function. The weights and biases constitute the model's parameters, which are optimized during the learning process. Note that for the first layer in the model, $\bm{x}_{n-1}=\bm{x}_0$, which is simply the initial model input, $\bm{x}$. With the exception of the output layer, the typical choice for the activation function is the rectified linear unit, or ReLU \cite{nair2010}, defined by $g(z)=$ max$(0,z)$, which outputs the maximum between the input and zero. A common variant is the leaky ReLU \cite{maas2013}, where negative inputs are not set to zero, but are instead multiplied by a small constant $\alpha$. In the output layer, the \textit{softmax} function (multi-class generalization of the logistic sigmoid, see for example \cite{goodfellow2016}) is used to produce a multinoulli distribution representing the probability that input $\bm{x}$ belongs to each of the $K$ different categories represented in the model. The category with the greatest probability is generally taken to be the classification, however specific threshold cuts for each category can also be used.

During the learning process, the model is presented with a large number of training examples where each input, $\bm{x}$, has a single human assigned categorical label, $y$, which is taken by the model to be the ground truth. The ground truth label, $y$, is then typically represented in a conditional probability distribution, $q$, such that the conditional probability for the $k$th category in the model is given by $q(k|\bm{x})=\delta_{ky}$, which is the Kronecker delta. A loss function is used to compute the error between the model predictions and the ground truth. Modern neural networks are typically trained using the principle of maximum likelihood. In this approach, the loss function is the negative log-likelihood, which can be equivalently described as the cross-entropy between the training examples and the modeled distributions \cite{goodfellow2016}. In the case of multinomial logistic regression (i.e., classification with multiple categories), the cross-entropy loss function for a single training example 
is:
\begin{equation}
H(p,q) = -\sum\limits_{k=1}^K q(k|\bm{x})\log(p(k|\bm{x})),\label{loss_func}
\end{equation}
where $K$ is the total number of categories in the model, $q(k|\bm{x})$ is the ground truth, human-assigned probability for the $k$th category, and $p(k|\bm{x})$ is the probability output by the model for the $k$th category. The gradient of the loss, as a function of the weights and biases, is calculated using the back-propagation algorithm \cite{rumelhart1986}. The loss is then minimized by updating the weights and biases for all the neurons in each layer using the method of mini-batch stochastic gradient descent (SGD)  \cite{lecun1998c,bottou2016}. When using mini-batches, the gradient of the loss function is estimated as the average instantaneous gradient over a small group of training examples (25 to 100, typically), which serves to balance gradient stability with computing time. This procedure is then repeated, iterating through mini-batches of training examples, until the error between the modeled and ground truth distributions reaches a satisfactory level. A single cycle through all of the mini-batches contained in the training set is typically referred to as an epoch.   

\subsection{Convolutional Neural Networks}\label{sec_cnn}
Convolutional neural networks (CNNs) \cite{lecun1998a} are a subclass of neural networks in which standard matrix multiplication is replaced with the convolution operation in at least one of the model's layers. CNNs have shown extraordinarily good performance learning features from datasets that are characterized by a known grid-like topology, such as pixels in an image or samples in a waveform. The core concept behind CNNs is to build many layers of ``feature detectors'' that take into account the topological and morphological structure of the input data \cite{simard2003}. Throughout the training process, the model learns how to extract meaningful features from the input, which can then be used to model the contents of the input data. The first stages of a CNN typically contain two types of alternating layers that are used to perform ``feature extraction'': convolutional layers and pooling layers. 
 
Convolutional layers take a stack of inputs (e.g. color channels in an image) and convolve each with a set of learnable filters to produce a stack of output feature maps, where each feature map is simply a filtered version of the input data (input image, in our case). A given input image, $I$, convolved with a $n\times m$ filter, $F$, will produce an output according to:
\begin{equation}
X_{p,q}=(F*I)_{p,q}=\sum\limits_{i=1}^n\sum\limits_{j=1}^m\sum\limits_{k=1}^cF_{i,j,k}\cdot I_{p+i,q+j,c},\label{conv_op}
\end{equation}
where $X_{p,q}$ is the $(p,q)$ pixel of the feature map (prior to applying the non-linear function), $n$ and $m$ correspond to the filter's height and width in units of pixels, and $c$ is the number of color channels in the input image\footnote{In our application, we sum the three color channels R, G, and B to produce a single grayscale color channel.}. With this transformation in mind, a slightly modified version of Equation~\ref{neuron}  can be constructed such that the $l$th of $L$ total feature maps output by the $n$th layer, $\bm{X}^{(l)}_n$, can be expressed with the following matrix representation:
\begin{equation}
\bm{X}_n^{(l)} = g\left(\sum\limits_{k=1}^K\bm{W}_n^{(k,l)}*\bm{X}_{n-1}^{(k)} +b_n^{(l)} \right),\label{feature_map}
\end{equation}
where $\bm{X}^{(k)}_{n-1}$ is the $k$th of $K$ total feature maps output by the previous layer\footnote{The input layer, $\bm{X}_{n-1}^{(k)}=\bm{X}_0$, isn't a feature map but is simply the input image for the model.}, $\bm{W}_n^{(k,l)}$ is a set of matrices containing the weights for the learnable filters, $b_n^{(l)}$ is the bias for the $l$th feature map, $*$ is the two-dimensional convolution operation shown in Equation~\ref{conv_op}, and $g$ is the activation function that performs a non-linear transformation of each pixel to produce the resulting feature map. 

Feature maps are essentially abstract representations of the input image, where each individual feature map is tasked with learning how to extract a specific feature from the input, such as edges, corners, contours, parts of objects, etc. It should be noted that the specific features learned by each feature map are not predetermined, but, rather, are selected solely by the model during the learning process. The feature maps nearest the input tend to resemble the original image. At layers further from the input, the feature maps gradually become more abstract and specialized.



Replacing the matrix product with a sum of convolutions results in a series of additional benefits \cite{goodfellow2016}: (1) a restricted connectivity pattern where each neuron is only connected to a local subset of the input, which reduces the number of computations, (2) the model learns a single set of parameters for each filter that can then be shared via convolution by all pixels in the input, which reduces the number of model parameters and improves the model's generalization performance\footnote{Generalization performance is a model's ability to perform well on previously unseen examples that were not included in the training set.}, and (3) the form of parameter sharing used in convolution also results in translation equivariance, meaning a translation in the input results in the same translation in the output. The restricted connectivity pattern results in the model learning predominantly from only local interactions in the input, meaning that features at distant locations of the input are less likely to interact. To combat this, convolutional layers are often used alongside pooling (subsampling) layers.

Pooling layers \cite{boureau2010} reduce the dimensionality of a feature map by using an aggregation function to compute a summary statistic across a small, local region of the input. The dimensional reduction gives the deeper layers of the model the ability to learn correlations between increasingly larger, yet lower resolution, regions of the input. For example, max pooling \cite{zhou1988} computes the maximum output located within a rectangular region of the input, then reduces that rectangular region to a single value equal to the maximum. A common choice is to divide each feature map into non-overlapping $2\times2$ grids of pixels that are then each reduced to a single pixel, converting a feature map from, say, $32\times32$ pixels to $16\times16$ pixels. As a result, only the most pronounced features in each rectangular region are forwarded to the deeper layers of the model. The pooling operation also gives rise to translation invariance\footnote{To be clear, $f$ is translation \textit{equivariant} if $f(T(x))=T(f(x))$, and translation \textit{invariant} if $f(T(x))=f(x)$, where $T(x)$ is a translation operation.} across small regions of the input. This is a desirable benefit when one is primarily interested in whether certain features are present in the input, rather than knowing precisely where they are located.

Finally, the features extracted from convolutional and pooling layers are typically used as input for a standard, fully connected, feedforward neural network (as explained in Section~\ref{sec_background}) where the desired output is then produced, which, in this case, is the CNN classification of the input image.

\section{Constructing a DECO CNN}\label{sec_deco_cnn}
In the sections that follow, we describe the construction and optimization of a DECO-specific convolutional neural network. We begin by introducing the dataset and the challenges associated with both human classification error and the small number of training images. We explain how data augmentation was used to make the model approximately invariant to rotations as well as artificially boost the number of training images. We then discuss the problem of overfitting and the techniques used to address it. Next, we summarize the model structure and training process used. Finally, we present the classification results, evaluate the performance of the model, and discuss the model's role in current and future DECO analyses. 

\subsection{Image Database and Human Labels}\label{sec_labels}
As discussed in Section~\ref{sec_background}, the model must not only be presented with a large number of training examples, but also with a set of corresponding human-determined categorical labels. However, assigning human labels to large datasets is time consuming and, depending on the dataset, difficult to do accurately. Previous deep learning models within the astronomy and particle physics communities have constructed labeled datasets by using a crowd-sourcing approach, for example by Galaxy Zoo~\cite{willett2013,dieleman2015}, or large-scale Monte Carlo event simulations, for example by the NOvA neutrino experiment~\cite{ayres2007,aurisano2016}. Both approaches require considerable human labor. At present, the DECO image database contains $\sim$45,000 events (images that passed the online filter), each of which potentially contains one or more clusters. Assigning human labels to each event cluster would be a very time consuming task. With this in mind, rather than labeling the entire dataset, we instead opted for an iterative approach in which the number of labeled training examples was successively increased in parallel with the optimization of the CNN model structure.

\begin{figure}[h]
  \centering
  \includegraphics[width=1\linewidth]{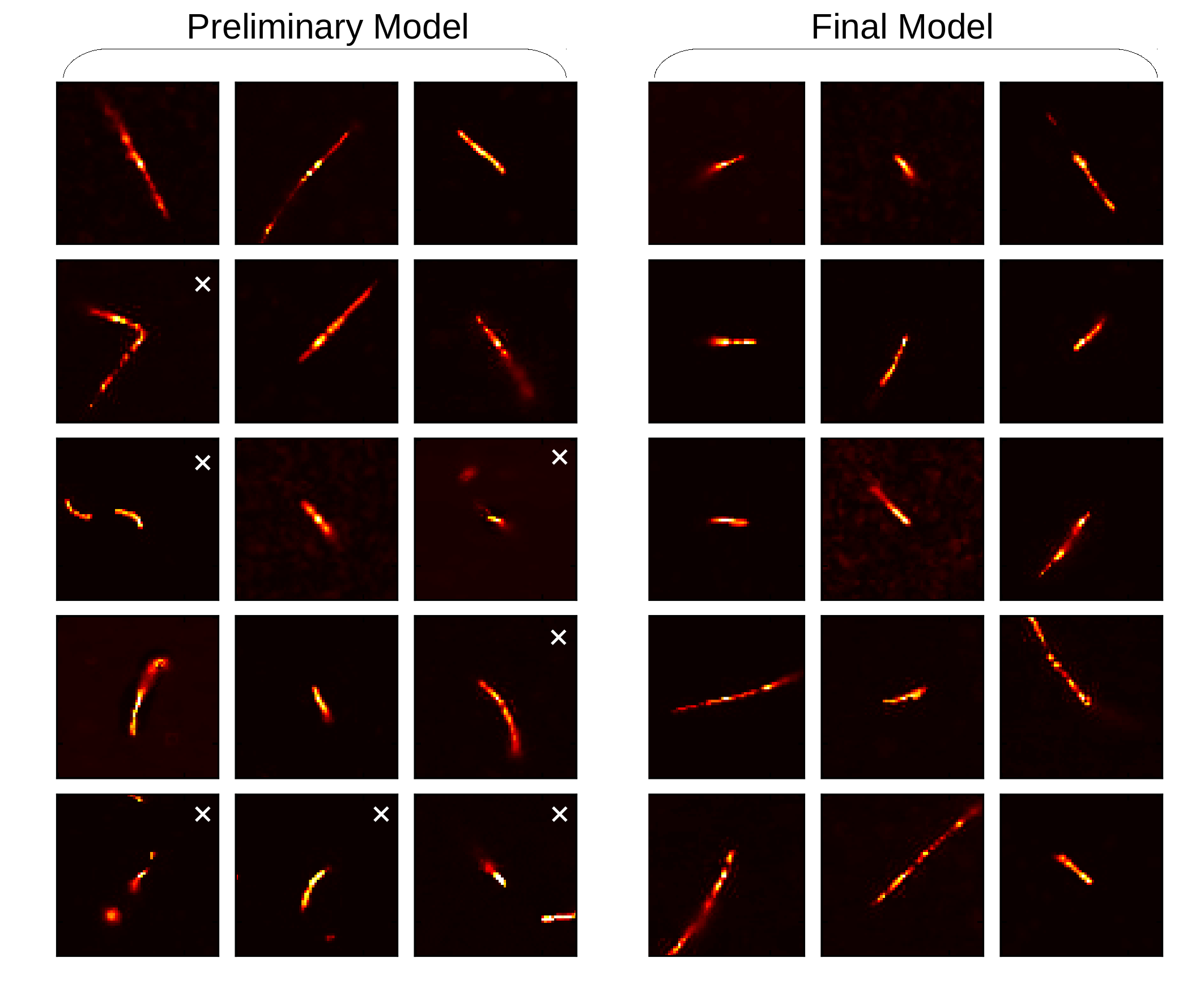}
  \caption{\small{Left: random sample of images with track probability $>$ 0.95 according to a preliminary version of the CNN model (presented in  \cite{meehan2017}). This version of the model struggled to correctly identify tracks that had similar features to other event types, particularly worms. Incorrectly classified images, denoted with a white `$\times$', were assigned a human label (worm, in each example shown) and added to the training set for the next iteration of the model. Right: random sample of images with track probability $>$ 0.95 according to the final version of the CNN model. The CNN classification agrees with the human classification for every single event in this sample.}}
  \label{model_grid}
\end{figure}

To accomplish this, individual event clusters were inspected by eye, by multiple people, and assigned labels of track, spot, worm, noise\footnote{The noise category was added during the iterative training process when it was found to drastically improve the model's overall classification accuracy.}, or ambiguous. Additionally, if a clear identification could not be made or if humans disagreed on the classification, which occurred $\sim$10$\%$ of the time, the image was labeled as ambiguous and excluded from the training set. During the optimization process, the model was trained and used to classify events that were not in the original training sample. These classified images were then searched by eye for likely false positives, i.e., instances where the model reports a high probability that an event belongs to a certain category but appears to be wrong. These incorrectly classified events were then assigned a correct human label, added to the existing set of training images, and used to train the next iteration of the model. This process was repeated on increasingly larger sets of images. As shown in Figure~\ref{model_grid}, with each new iteration, the examples that the model found most difficult to categorize were added to the labeled dataset, thus addressing the remaining weaknesses in the classifier.

\begin{figure}[h]
  \centering
  \includegraphics[width=1\linewidth]{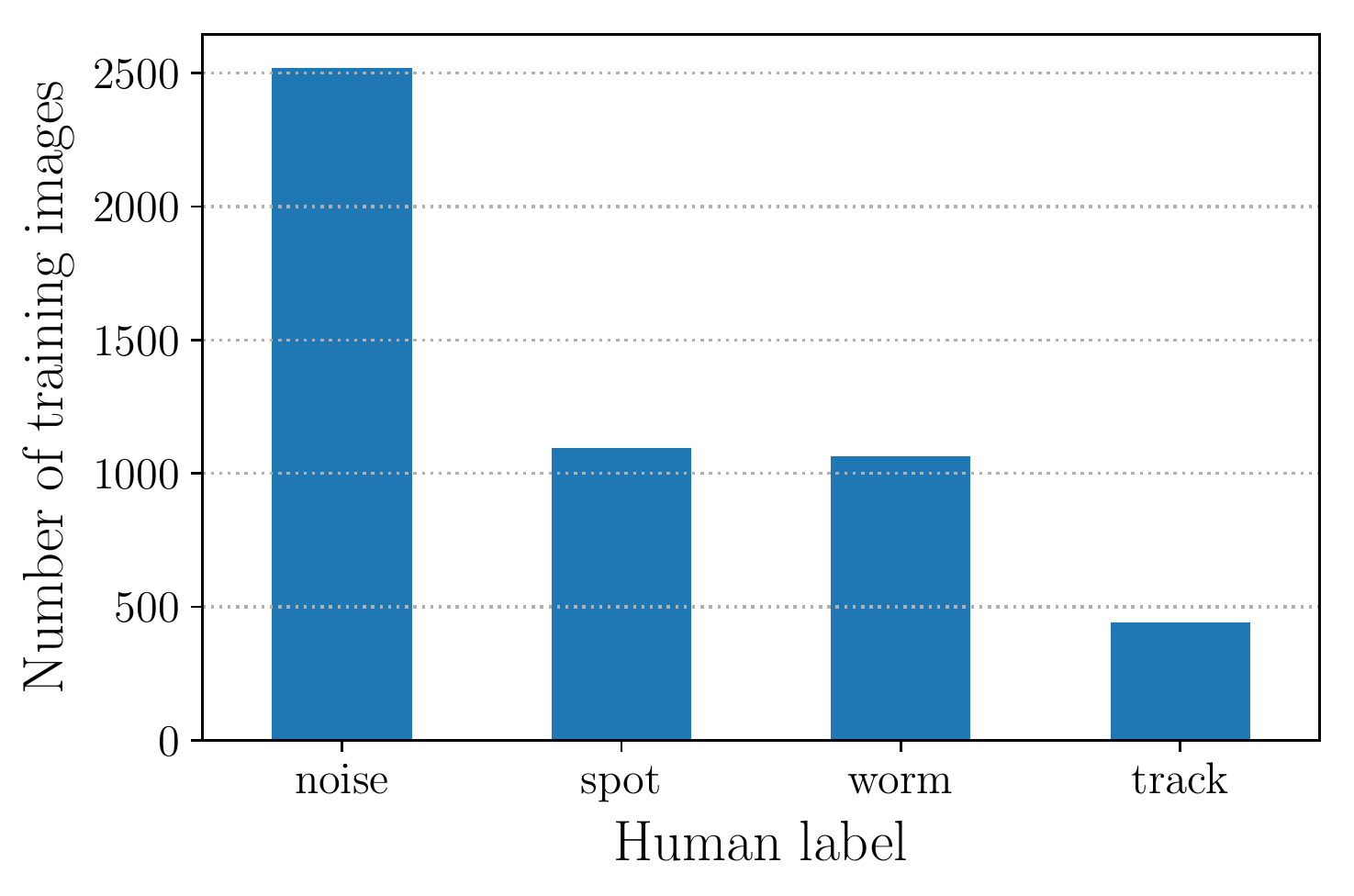}
  \caption{\small{Number of training images for each event type contained in the final dataset that was used to train the best performing model. Out of the 5119 total images, there are 2520 (49\%) noise, 1094 (21\%) spot, 1063 (21\%) worm, and 442 (9\%) track images.}}
  \label{samples}
\end{figure}

\subsection{Preprocessing and Data Augmentation}\label{sec_data_aug}
Image-to-image variations in position, scale, and rotation pose a challenge to DECO event classification. When a DECO user collects data, both the position and orientation (at least in azimuth -- zenith typically corresponds to phones operating flat on a table) of the phone is arbitrary. Both orientation and location data are collected in the app's metadata. However, the $(x,y)$ position of a given event cluster within the camera sensor, as far as the model is concerned, should be considered a meaningless feature. Similarly, the orientation of a hit cluster within the $(x,y)$ plane, as well as reasonable variations in scale (e.g. the length of a track) should also be considered meaningless by the model. Fortunately, CNNs naturally handle translations in the input quite well \cite{lecun1998b,gong2014}. However, invariance to features such as scale and rotation need to be learned.

For a given input image, the apparent size of the event with respect to the camera sensor can be affected by a number of factors such as the underlying hardware in the specific phone model (including the image sensor resolution), the energy of the particle, and the angle of incidence. The pooling operation provides resiliency to minor changes in shape and scale \cite{scherer2010}, however, variations larger than a few pixels must be addressed by other means. Sophisticated solutions to this problem have been proposed~\cite{xu2014}, however, the simplest method is to introduce scale-jittering via data augmentation, which is in widespread practice today \cite{simonyan2014,krizhevsky2012}. Data augmentation consists of randomly transforming training images while preserving their human-assigned category labels. Similar to scale invariance, data augmentation can also be used to learn rotation invariance. While rotation-invariant CNN architectures exist \cite{dieleman2015} and have been shown to outperform data augmentation in certain cases \cite{marcos2016}, the small number of training images in this study prohibited the use of such methods. Finally, due to the limited number of training images available, data augmentation was also used to artificially inflate the number of ``unique'' images seen by the model during training. 

In general, data augmentation has been shown to be the simplest way to achieve approximate invariance to a given set of transformations \cite{simard2003}. Assuming the model has the capacity to do so (i.e., enough feature maps), the model should be able to learn a wide variety of invariances directly from the data \cite{lenc2014}. An additional benefit of data augmentation is that a single set of transformations can be used to address multiple different issues. With that in mind, the following operations were applied to each training image:
\begin{itemize}
\itemsep0em 
\item \textbf{grayscale conversion and normalization}: a dimensional reduction over the channel axis of each image was performed by calculating an unweighted sum of each pixel's R+G+B value. The resulting grayscale images were then normalized to 1, taking the maximum possible R+G+B value to be 765 (i.e., $255 \times 3$). Grayscale reduces the variation seen from phone to phone and is also computationally more efficient. Furthermore, while color provides essential information for other image classification tasks, it does not for particle tracks.
\item \textbf{translation}: random left/right and up/down shifts, each by an integer number of pixels uniformly sampled between -8 and +8 with respect to the image center. 
\item \textbf{rescale}: random zoom in/out uniformly sampled between 90\% and 110\% of the original image size, used for learning scaling invariance.
\item \textbf{reflection}: random horizontal and vertical reflections, each with a probability of 50\%.
\item \textbf{rotation}: random rotation uniformly sampled between $0^\circ$ and $360^\circ$; used for learning rotation invariance. After the rotation, any remaining pixels outside the boundaries of the original input were assigned a value of 0. 
\item \textbf{crop}: crop from $100\times 100$ pixels to $64\times 64$ pixels; used to reduce the amount of empty space created on the boundaries of the image as a result of rotation, translation, and rescaling. 
\end{itemize}
With the exception of normalization and the conversion to grayscale, which could be performed ahead of time, all data augmentation was done in real time during the training process. Prior to the start of each training epoch (full cycle through all training images, as defined in Section~\ref{sec_background}), a new random set of perturbations are applied to each image. Applying data augmentation in this way ensures that the model is never presented with the exact same version of a training example more than once. Real-time data augmentation is performed in Python using the Keras neural network application programming interface \cite{keras}, which makes use of tools contained within the SciPy library \cite{scipy}.

\subsection{Avoiding Overfitting Through Regularization}\label{sec_overfitting}
Deep neural networks typically have anywhere from tens of thousands to tens of millions of trainable parameters. The advantage of such a large number of parameters is that the model has the ability to fit extremely complex and diverse datasets. However, the downside of a model with such tremendous freedom is that there is considerable risk of over-fitting, which occurs when the model simply memorizes the training images. As a result, the model is overly sensitive to the specific features that were memorized during training and therefore generalizes poorly to new data. Over-fitting is of particular concern when dealing with a small number of training images, as is the case in this study. To combat this phenomenon, we used several regularization techniques \cite{goodfellow2016,kukacka2017}, which are modifications to the learning process that are intended to reduce generalization error while leaving training error\footnote{The error between the true and predicted classification for images in the training set} unaffected. These techniques are as follows:
\begin{itemize}
\itemsep0em 
\item \textbf{data augmentation}: artificially increasing the number of training examples by modifying the images in such a way that they look different for each particular training instance while still maintaining the correctness of the underlying human assigned label. The particular perturbations used are outline in Section~\ref{sec_data_aug}.
\item \textbf{label smoothing}: accounting for the uncertainty in human assigned labels by replacing the hard 0, 1 (false, true) label distribution,  $q(k|\bm{x})=\delta_{ky}$, with $q(k|\bm{x})=(1-\epsilon)\delta_{ky}+\frac{\epsilon}{K}$, where $k$ is the $k$th of $K$ total categories in the model, $\epsilon$ is a small constant representing the probability of an incorrect label, and $y$ is the human label. This modification results in an additional penalty term being introduced into the loss function, Equation~\ref{loss_func}. Assuming that $\epsilon$ is reasonably small, this technique reduces the effect of incorrect labels while still encouraging correct classification  \cite{szegedy2015}.  
\item \textbf{dropout}: at every step of the training process, each individual neuron in a given layer has a probability, $P$, of being temporarily set to zero, or ``dropped out'' \cite{hinton2012,srivastava2014}. The purpose of dropout is to prevent the co-adaptation of neuron outputs such that each individual neuron depends less on other neurons being present in the network. To preserve the total scale of inputs, the neurons that weren't dropped out are rescaled by a factor of $1/(1-P)$. Dropout is only applied during training and turned off afterwards.
\item \textbf{max-norm constraint}: to prevent weights from blowing up, a max-norm constraint is applied to each neuron's weight vector, $\bm{W}$, such that $\Vert\bm{W}\Vert\leq r$, where $\Vert\cdot\Vert$ is the $L^2$ vector norm and $r$ is a user specified constant dictating the maximum value. After each training step the constraint is checked and, when necessary, the weights are updated according to $\bm{W}\rightarrow\bm{W}\frac{r}{\Vert\bm{W}\Vert}$. The max--norm constraint, both with and without dropout, has been shown to help reduce over-fitting \cite{srivastava2014,srebro2005}. This constraint was applied to fully connected layers only.
\item \textbf{early stopping}: during the training process, testing loss (error) typically decreases, reaches a minimum value, and then begins to increase again once over-fitting has set in. To avoid using an overfit model, we capture running snapshots of the best version of the model during training, which correspond to the epochs where testing loss reaches a new minimum value \cite{bishop1995,sjoberg1995}. 
\item \textbf{categorical weights}: As seen in Figure~\ref{samples}, certain event types, tracks in particular, have fewer training images than others. As a result, the model sees more training examples from the abundant categories than the under-represented ones, which introduces bias into the classifier. To account for this imbalance, each category is assigned a weight, according to its abundance, which is applied to the loss function (Equation~\ref{loss_func}) to ensure that all categories are represented equally during optimization.
\end{itemize}

\subsection{Model Structure and Training}\label{sec_training}
\begin{figure*}[h!]
  \centering
  \includegraphics[width=1\linewidth]{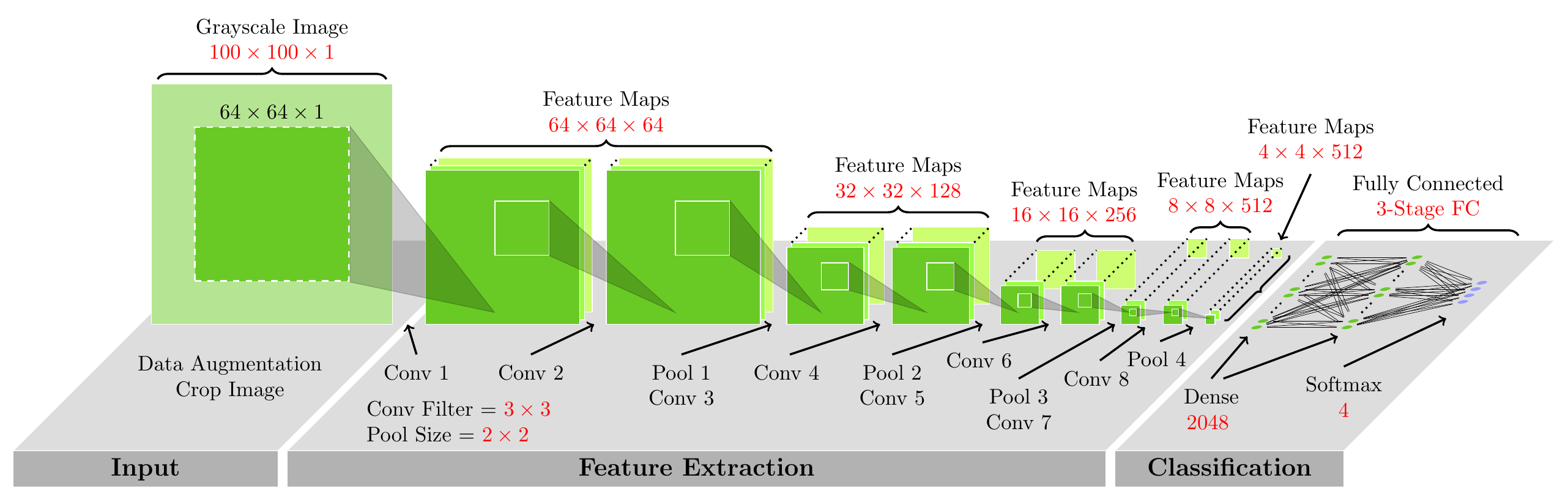}
  \caption{\small{Block diagram of the best performing network trained in this study. The input and output dimensions for each operation are shown to the left and right of the arrows, respectively. All convolutional filters are $3\times3$ and all pooling operations are $2\times2$ max pooling. Following the fourth pooling layer, the feature maps are flattened to a single 1-dimensional vector of length 8196, which is then used as input for the first dense layer.}}
  \label{cnn_diagram}
\end{figure*}

The best performing model trained in this study begins by taking a normalized, $100\times100$ grayscale image (zoomed in on the hit pixel cluster) as input. The input is then transformed via data augmentation (Section~\ref{sec_data_aug}), cropped to $64\times64$, and subjected to dropout with a probability $P=0.2$. Next, feature extraction is performed using four three-layer-deep blocks, each of which consists of the following operations: $3\times3$ convolution followed by a leaky ReLU activation, a second identical $3\times3$ convolution with leaky ReLU, and, lastly, $2\times2$ max pooling. For the leaky ReLU non-linearity, a constant multiplier $\alpha=0.3$ is applied for all negative inputs. Following max pooling in each block, dropout is applied with probability $P=0.2$. For each of the four blocks, the number of feature maps is doubled, starting with 64 in the first block and ending with 512 in the last. The model structure is loosely based on the VGG-16 network \cite{simonyan2014}, which used only $3\times3$ convolutional filters and $2\times2$ max-pooling throughout the network. Following feature extraction, the feature maps are flattened to a single, one-dimensional vector that is used as input for a three-layer fully connected network (Section~\ref{sec_background}). The first two layers are identical dense (fully connected) layers with 2048 neurons, leaky ReLU activation with $\alpha=0.3$, and a max-norm constraint with $r=3$ (see Section~\ref{sec_overfitting}). Each dense layer is also followed by dropout with a probability $P=0.4$. Finally, the output layer performs softmax regression, which outputs the probability for each of the 4 categories in the model (track, spot, worm, and noise). Figure~\ref{cnn_diagram} shows a block diagram of the model structure and workflow. Specific details for each layer are summarized in Table~\ref{tbl_cnn}. 

To train the model, we used a variant of mini-batch SGD (see Section~\ref{sec_cnn}) known as Adadelta~\cite{zeiler2012}. For our model, Adadelta was found to converge slightly faster than both SGD and Adam~\cite{kingma2014}, another widely used variant of SGD. At the beginning of each training epoch, a new set of random data augmentation perturbations are applied to each image in the training set. The model was programmed in Python using the Keras neural network application programming interface \cite{keras} operating with a Theano \cite{theano} backend. The final model contains approximately 25 million trainable parameters and was trained on a single NVIDIA Quadro M4000 graphics processing unit (GPU) with 8~GB of RAM.

\begin{table}[h]
\small
\centering
\begin{tabular}{c c c c c c c}
\toprule
 &\bf{Layer}&\bf{Features}&\bf{Size}&\bf{Activation}&\bf{Dropout}\\[.5ex]
\hline
\bf{1} & Convolution	&	64	&	$3\times3$	&	Leaky ReLU	&	-	\\
\bf{2} & Convolution	&	64	&	$3\times3$	&	Leaky ReLU	&	-	\\
\bf{3} & Max Pooling	&	-	&	$2\times2$	&	-	&	0.2	\\

\bf{4} & Convolution	&	128	&	$3\times3$	&	Leaky ReLU	&	-	\\
\bf{5} & Convolution	&	128	&	$3\times3$	&	Leaky ReLU	&	-	\\
\bf{6} & Max Pooling	&	-	&	$2\times2$	&	-	&	0.2	\\

\bf{7} & Convolution	&	256	&	$3\times3$	&	Leaky ReLU	&	-	\\
\bf{8} & Convolution	&	256	&	$3\times3$	&	Leaky ReLU	&	-	\\
\bf{9} & Max Pooling	&	-	&	$2\times2$	&	-	&	0.2	\\

\bf{10} & Convolution	&	512	&	$3\times3$	&	Leaky ReLU	&	-	\\
\bf{11} & Convolution	&	512	&	$3\times3$	&	Leaky ReLU	&	-	\\
\bf{12} & Max Pooling	&	-	&	$2\times2$	&	-	&	0.2	\\

\bf{9} & Dense 	&	2048	&	-	&	Leaky ReLU	&	0.4 \\
\bf{10} & Dense	&	2048	&	-	&	Leaky ReLU	&	0.4	\\
\bf{11} & Dense	&	4	&	-	&	softmax	&	-	\\[1ex]
\bottomrule
\end{tabular}
\caption{\small{Layer-by-layer summary of the best performing network. Each layer name is given followed by the number of feature maps (convolutional layers) or neurons (dense layers), the size of the convolutional filter or pooling region, the activation function used, and, lastly, the amount of dropout applied. For the leaky ReLU activation function, the value of $\alpha$ was set to $0.3$ in all cases. A max-norm constraint of 3 was used for both 2048 dense (fully connected) layers. Dropout with a probability $P=0.2$ was also applied to the input layer (not listed in the table).}}
\label{tbl_cnn}
\end{table}

\section{Results and Analysis}\label{sec_results}
\subsection{Model Performance}\label{sec_performance}
\begin{figure}[h!]
  \centering
  \includegraphics[width=1\linewidth]{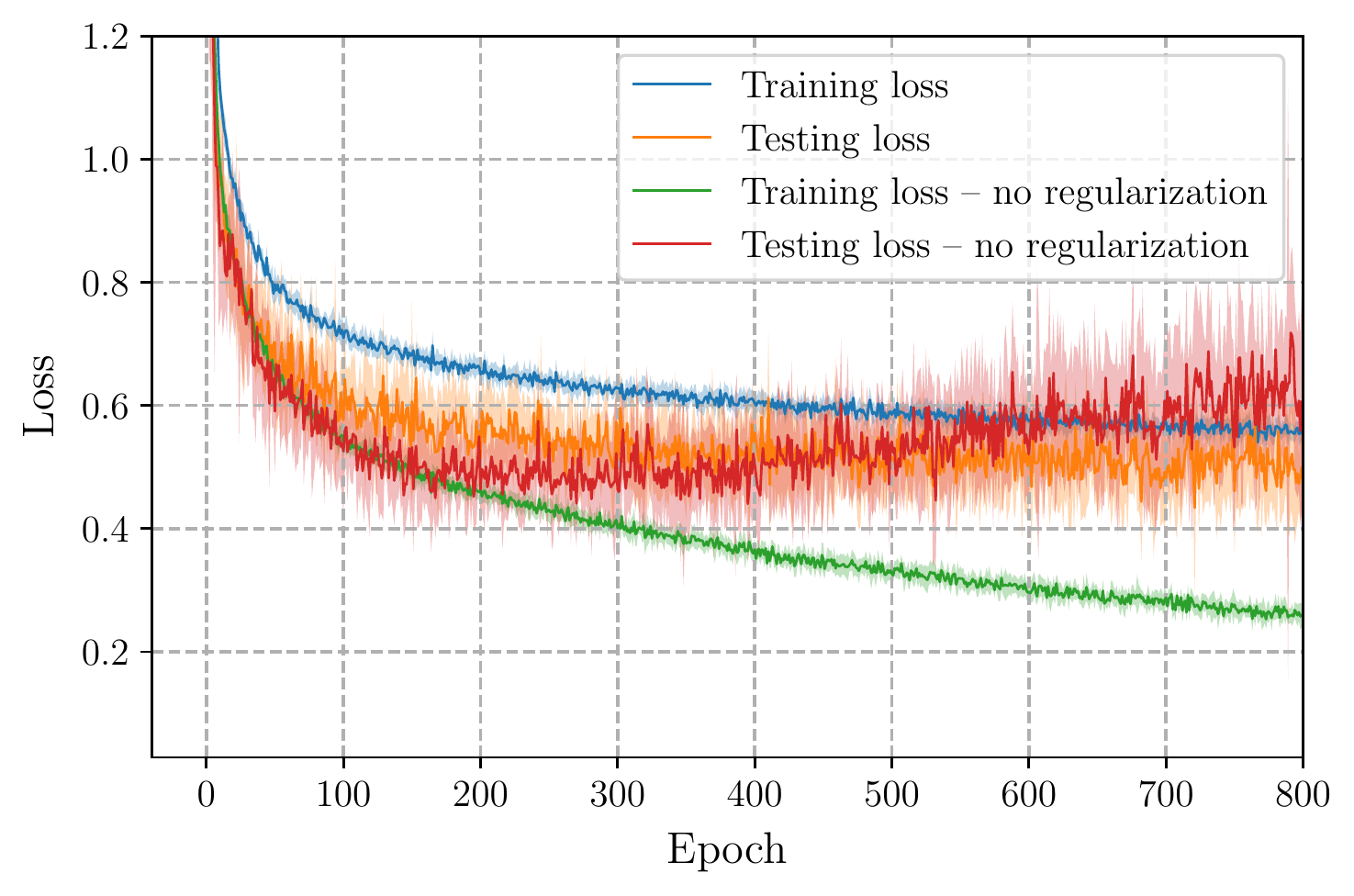}
  \caption{\small{Loss as a function of epoch for two different versions of the model, one trained with regularization techniques and one trained without. The loss is averaged over the 10-fold cross validation of the entire dataset and shaded error bands indicate the $1\sigma$ spread across the 10-folds. An epoch refers to one full cycle through all available training images.}}
  \label{fig_loss}
\end{figure}

To estimate the overall performance of the model, independent sets of human-classified images were evaluated using the method of stratified k-fold cross-validation \cite{kohavi1995}. In this procedure, the set of training images is split into $k$ groups, where each group contains a roughly equal number of images from each of the categories represented in the model. $k$ otherwise identical versions of the model are then trained, each time setting aside one group for testing and $k-1$ for training the model. Selecting a value of 10 for $k$, we trained each individual fold for a total of 800 epochs, where each epoch consists of a single cycle through the full set of training images. The loss (defined below) for both training and testing sets, averaged over the 10 folds as a function of training epoch, is shown in Figure~\ref{fig_loss}. The loss\footnote{Note that this is technically the logarithm of the loss and therefore is not expressed as a percentage.} for a set of examples is defined to be:
\begin{equation}
\mathcal{L}=-\frac{1}{N}\sum\limits_{n=1}^N\sum\limits_{k=1}^K q_n(k|\bm{x})\log(p_n(k|\bm{x}))w_k,
\end{equation}
where $N$ is the number of training or testing images, $K=4$ is the number of categories in the model, $p$ and $q$ are the respective CNN and human assigned categorical distributions for each image (defined in Section~\ref{sec_background}), and $w$ is a categorical weight term to account for the categorical imbalance in the training set (see Section~\ref{sec_data_aug}). Conceptually, the loss can be thought of as the average error between the human and CNN classifications. 

    Early stopping (Section~\ref{sec_overfitting}) was used to obtain the best performing (lowest testing loss) versions of the model throughout each 800-epoch training session, which, on average, occurred near epoch 650. The training and testing loss as a function of training epoch can be seen in Figure~\ref{fig_loss}. The gap between the training and testing loss is caused by the regularization techniques used to prevent overfitting, which are only applied to the training set (see Section~\ref{sec_overfitting}). Lower testing loss than training loss can also be indicative of an underfit model. To test this, an alternate version of the model was trained with dropout removed from all layers, the max-norm constraint removed from the fully-connected layers, and no label smoothing. The results of this test revealed that the gap between testing and training loss disappeared until overfitting set in at epoch $\sim$200. This explains the gap between training and testing loss and also confirms that the regularization techniques are effectively preventing the model from overfitting the data. To investigate the potential benefits of a longer training duration, an additional model was trained for 10,000 epochs. While training loss was observed to decrease slightly, no benefit was seen in the testing set, thus confirming that 800 epochs was sufficient. A value of $\epsilon=0.004$ was used for label smoothing. Setting $\epsilon$ to 0 as well as using larger values of $0.1$ and $0.01$ all resulted in marginally higher testing loss. We also tested an alternate, simpler version of the model which is described in Section~\ref{sec_simple_model}.

\subsection{Model Accuracy}\label{sec_accuracy}
\begin{figure}[h]
  \centering
  \includegraphics[width=1\linewidth]{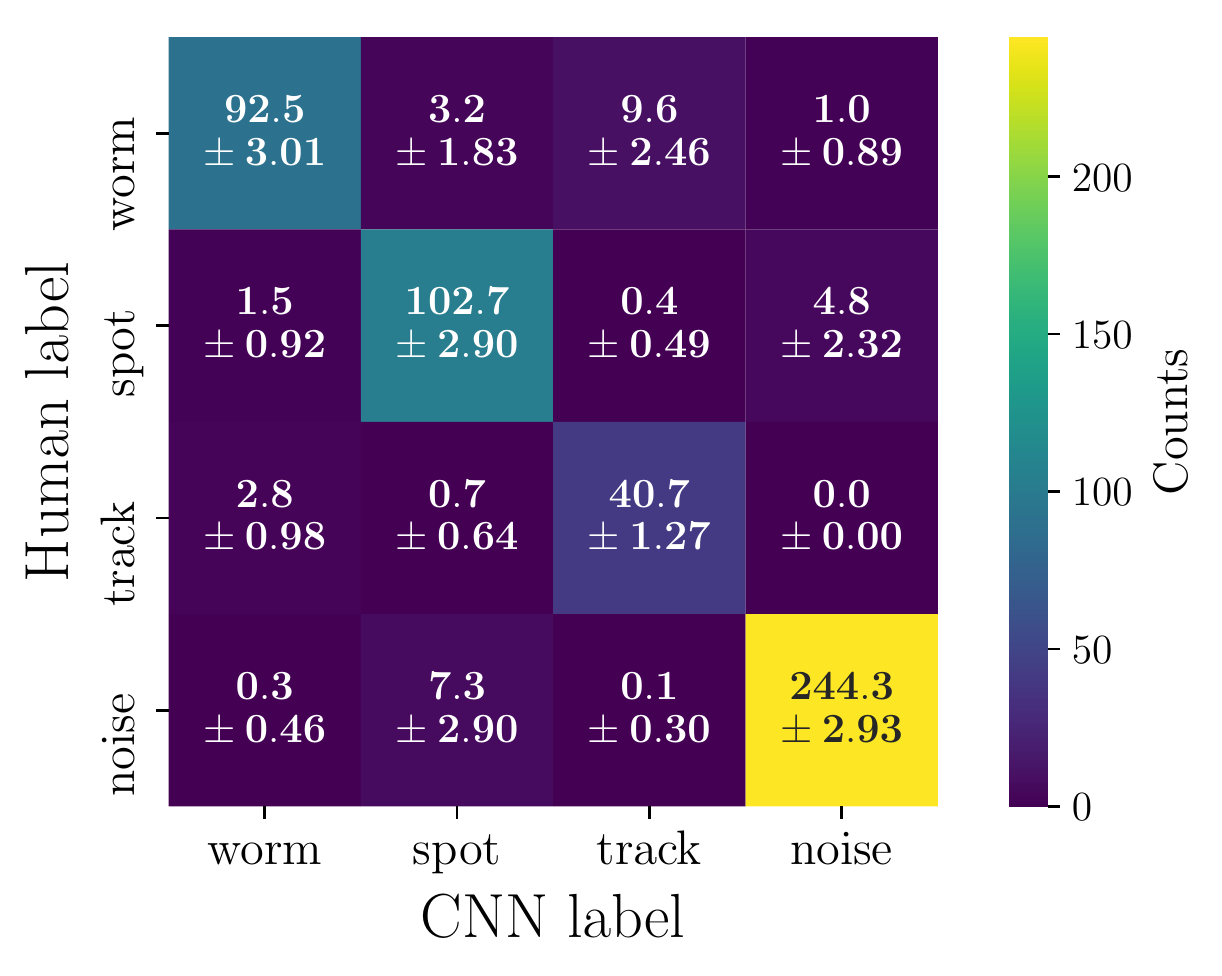}
  \caption{\small{Confusion matrix summarizing the CNN categorization accuracy. The vertical axis shows the ground truth (human-determined) classification and the horizontal axis shows the classification from the CNN. The values shown in the confusion matrix are the average and standard deviation of the testing set results from the 10-fold cross validation.}}
  \label{conf_mat_counts}
\end{figure}

\begin{figure}[h]
  \centering
  \includegraphics[width=1\linewidth]{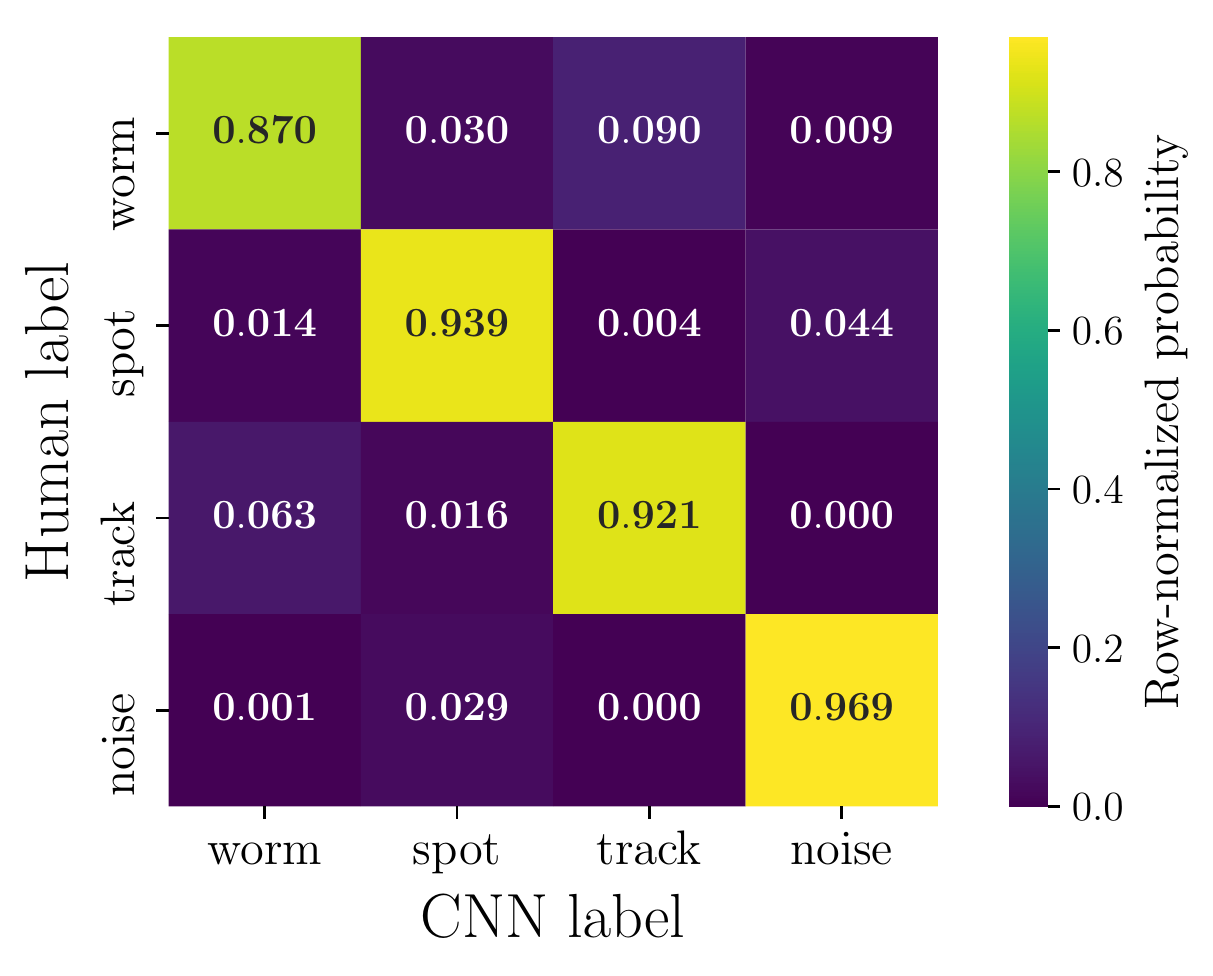}
  \caption{\small{Row-normalized confusion matrix that accounts for the relative imbalance in the number of testing examples for each category in the training set. Normalization is performed independently for each row and is calculated by dividing each row of the unnormalized confusion matrix (see Figure 10) by the total number of events in that row.}}
  \label{conf_mat_rows}
\end{figure}

Figure~\ref{conf_mat_counts} shows a category-by-category summary, known as a confusion matrix, quantifying the error between human and CNN classifications for each category in the model. Each square of this confusion matrix is calculated by averaging the testing set results over the 10 folds in the cross validation. It should be noted that the resulting distribution is not normalized and is biased according to the relative occurrence of each category in the training set. For example, noise events make up almost half of the training set (Figure ~\ref{samples}). This bias can be removed by normalizing each row of the confusion matrix to the total counts contained in each row, i.e. the total number of human-labeled events for each category. The resulting row-normalized confusion matrix describes the conditional CNN probability distributions for each of the four human-assigned labels in the model. The probability of the CNN correctly identifying each event type, along with the probability of mis-identifying each category, can be read directly off of the row-normalized confusion matrix in Figure~\ref{conf_mat_rows}. For example, the model correctly identifies human-labeled tracks as tracks 92\% of the time, while incorrectly identifying them as worms 9\% of the time. This confusion in the classifier is both expected and comparable to human performance, given that, out of the four categories in the model, track and worm event morphologies are among the most similar. The model accurately labels noise events 97\% of the time, which is the highest accuracy of any event type. This is also expected due to the vast differences between charged particle events and noise. Moreover, this also confirms that the model successfully learned the concept of noise, justifying the inclusion of this category in the model.

These results assume that a single classification is assigned to each image by choosing the category with the highest CNN output probability. We explore the performance of alternative choices below.

\begin{figure}[h]
  \centering
  \includegraphics[width=1\linewidth]{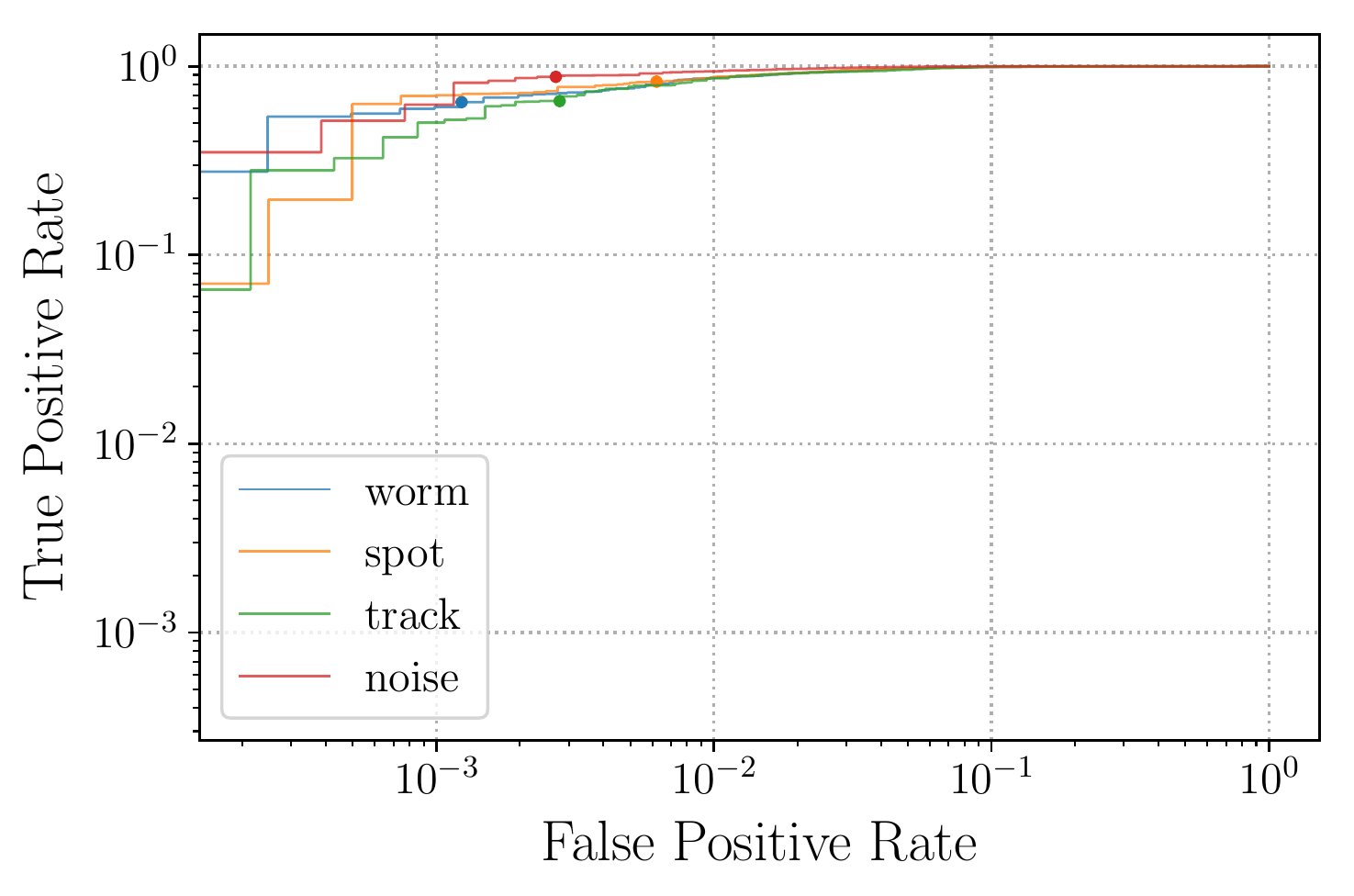}\\
  ~~\includegraphics[width=1\linewidth]{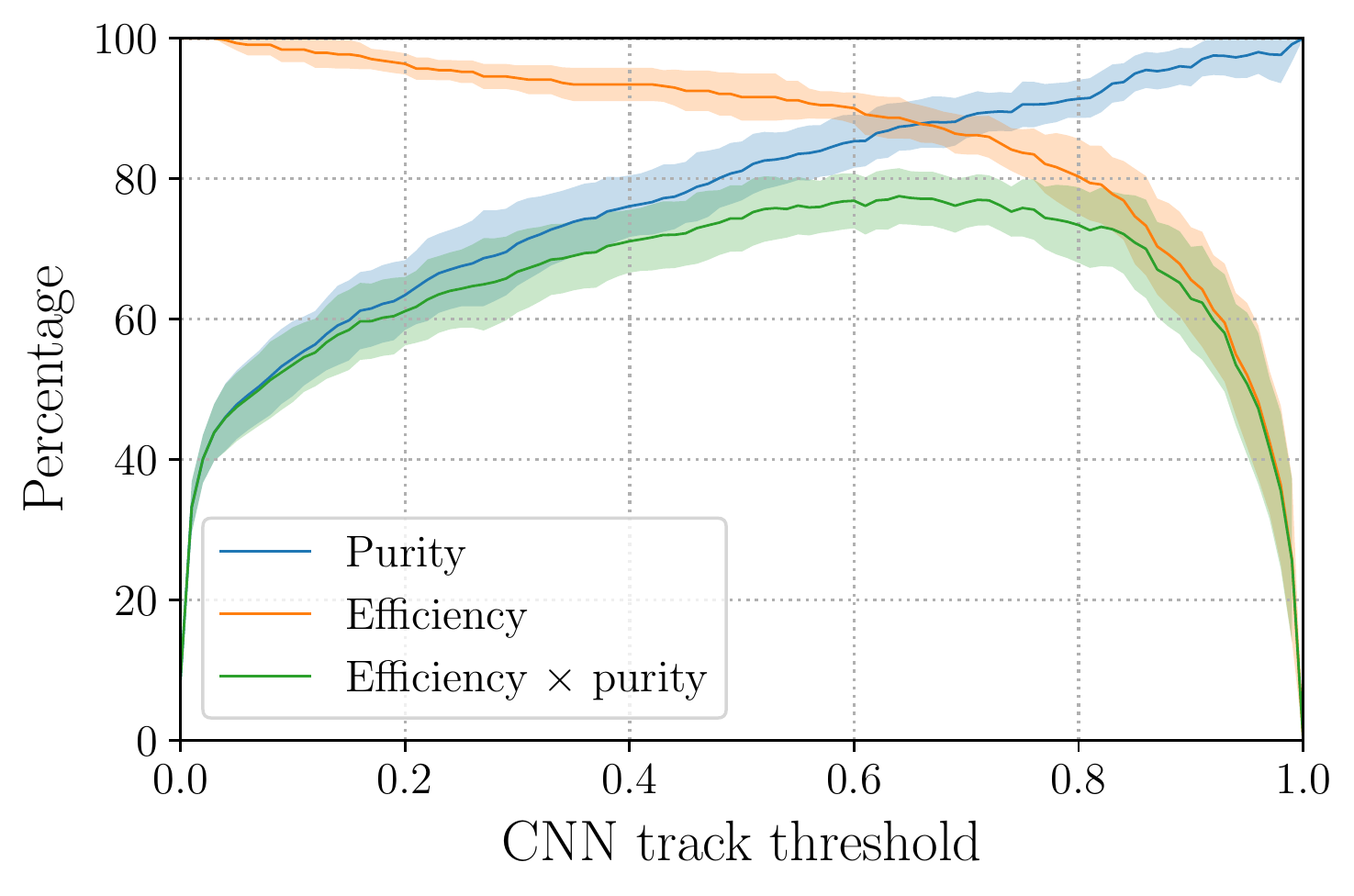}
  \caption{\small{(Top) Receiver operating characteristic (ROC) curve displaying the true positive rate vs. false positive rate for a variety of threshold values. A threshold of 0.9 is indicated with a dot for each category. (Bottom) Purity, efficiency, and their product as a function of CNN track probability threshold, averaged over the 10-fold cross validation. For each curve, the average and standard deviation are indicated by the thin solid line and corresponding band, respectively. For each threshold value, the purity and efficiency are calculated for events with a CNN track output, $p_{track}$, above the track threshold.}}
  \label{fig_roc_purity}
\end{figure}

We further evaluate the model's classification performance by calculating the true and false positive rates for each category, assuming a binary classification scheme (e.g. track and non-track). The true and false positive rates for each category are parameterized according to a threshold applied to its CNN output probability and plotted as a receiver operating characteristic (ROC) curve, as seen in the top panel of Figure~\ref{fig_roc_purity}. For example, requiring a track probability of at least 0.9 results in a true positive rate of 60\% and a false positive rate of 0.3\%. While the trade-off between efficiency and purity\footnote{The definitions of purity and efficiency used here are generally referred to as precision and recall, respectively, within the machine learning community.} can be inferred from the ROC curve, these quantities were also explicitly calculated for tracks, which is the primary category of interest for most DECO users. The resulting efficiency, purity, and efficiency $\times$ purity curves, averaged over the 10 folds and plotted as a function of track probability threshold, are shown in the bottom panel of Figure~\ref{fig_roc_purity}. For a given fold and threshold, the efficiency is calculated from the testing set and defined to be the ratio of the number of tracks that pass the threshold to the total number of tracks. Likewise, for a given fold and corresponding test set, the purity is defined as the ratio of the number of human-labeled tracks that pass the threshold to the total number of events, regardless of event type, that pass the threshold. The product of the resulting curves is one metric that can be used to determine a threshold value that balances the efficiency vs. purity trade-off. 

\subsection{Comparison With Simpler Model}\label{sec_simple_model}
In the previous sections, we have shown that the model exhibits excellent performance across all four categories when classifying unseen data. However, one might wonder if the complexity of our model, which contains 25 million trainable parameters, is necessary to achieve this level of performance. In order to test this, we trained a simpler version of the model, containing 140 thousand parameters, with the same efforts described in Sections ~\ref{sec_data_aug} and ~\ref{sec_overfitting}. The simpler model contained only two blocks of convolutional and pooling layers, followed by significantly smaller dense layers than those described in Section ~\ref{sec_training}. The performance of this model was evaluated using the same 10-fold cross-validation described in Section ~\ref{sec_performance}. Compared to our more complex model, the simple model was equally accurate when classifying spots and noise, but 17\% less accurate at classifying worms and 7\% less accurate at classifying tracks. Furthermore, when evaluating the track performance in a binary fashion (see Section~\ref{sec_accuracy}), a 0.8 track threshold cut with the simple model resulted in a track sample with $<$80\% purity and only 40\% efficiency. This suggests that a more complex model is necessary in order to distinguish tracks and worms, which are the most interesting events scientifically. 

\subsection{Comparison With Straight Cuts}\label{sec_comp}
Early classification attempts, described in Section~\ref{sec_init_class}, sought to separate tracks from non-tracks in a binary fashion using straight cuts on simple metrics. This method, which used each image's area, number of clusters, and eccentricity, can be directly compared to the CNN model. To accomplish this, we treat the CNN output as a binary classification scheme (track or non-track) and evaluate both classification methods on the same set of testing images and corresponding human-assigned labels. The initial, straight-cuts model yielded a track selection with an efficiency of 69\% and a purity of 37\%. The low purity is likely due to small differences in the event topologies of many tracks and worms, which can be difficult to capture with simple geometric metrics. Moreover, optimization of the straight-cuts approach required aggressive cuts on these metrics, which also contributes to its poor efficiency in identifying tracks. The CNN classification, on the other hand, identifies tracks with 80\% efficiency and 91\% purity (cutting at a track probability threshold of 0.8, to be explained in Section~\ref{sec_dataset}), and can also accurately identify worms, spots, and noise with similar performance. Furthermore, the output probabilities of the CNN model enable us to design an event selection with a desired efficiency and/or purity in mind. 

\subsection{Application To Full Dataset}\label{sec_dataset}
While the CNN model has a number of uses, providing real-time classifications for the events listed in the public DECO data browser \cite{deco_data} is perhaps the most important. For this purpose, we seek to maintain a high-purity set of events identified as tracks. After evaluating constant cut-off values of 0.7, 0.8, and 0.9 on the testing set, we opted for a probability threshold of 0.8, which yields an event selection with a track efficiency of 80\% and, most importantly, a track purity of 91\%. As a result of applying a threshold cut rather than the maximum-probability criterion, there are some events with probability below threshold for every single category, which are therefore assigned a label of ``ambiguous''. More aggressive threshold cuts result in more events being labeled ``ambiguous''.

\begin{figure}[h]
  \centering
  \includegraphics[width=1\linewidth]{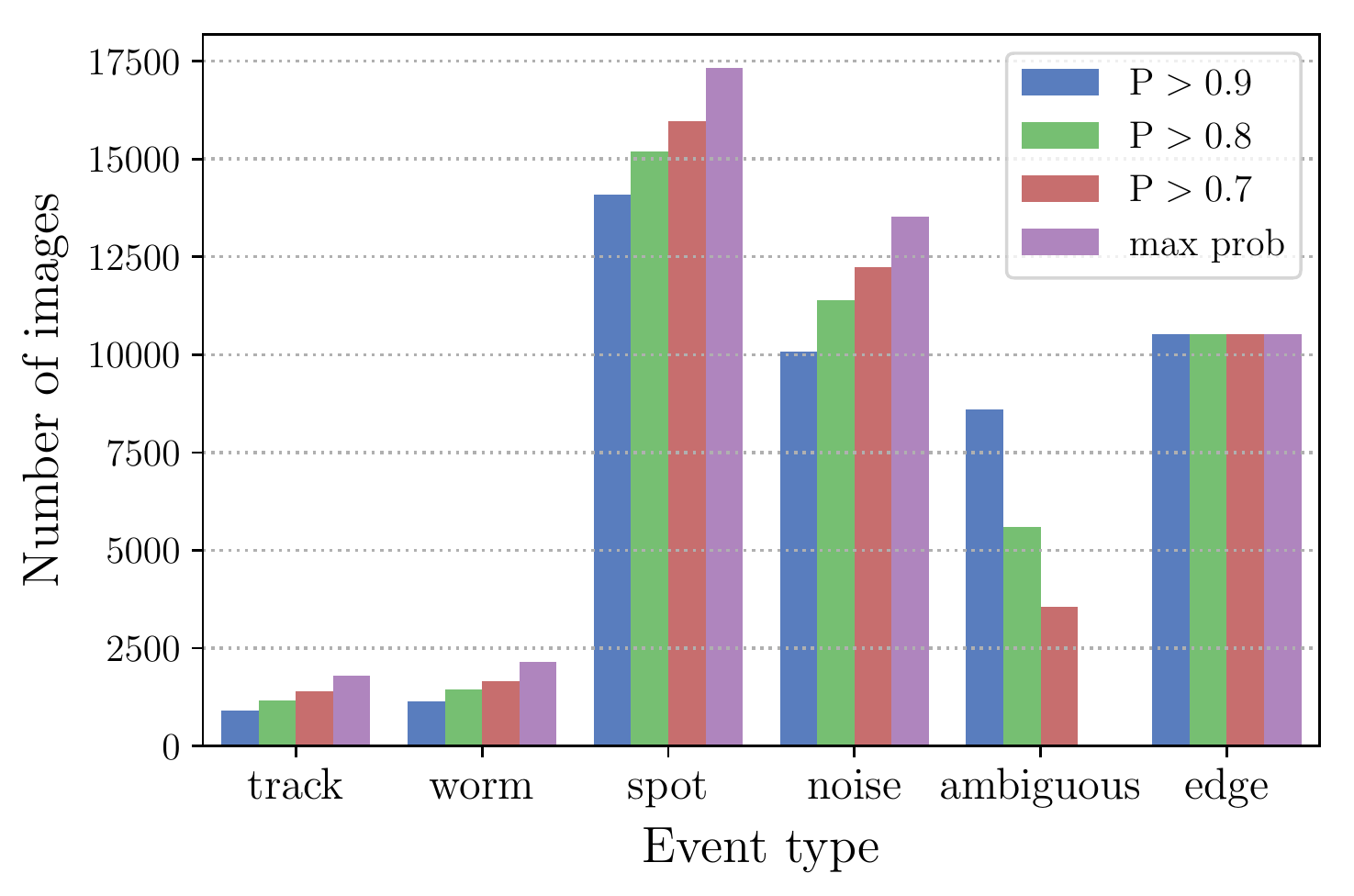}
  \caption{\small{Distribution of event types in the full data set (45,316 images) for different threshold choices applied to the CNN output probabilities. Each threshold is applied uniformly to all four categories and any event that does not have a probability greater than the threshold for any category is labeled ``ambiguous''. The fourth selection classifies events according to their maximum probability, which is why there are no ambiguous events in that scheme. ``Edge'' events are images with event clusters located less than 32 pixels from the camera sensor edge, which is incompatible with the CNN input requirement of 64$\times$64 pixel images. The relatively high rate of edge-type images may be due to light leakage around the edges of the image sensor when DECO is run under sub-optimal data-taking conditions, such as in a well-lit room.}}
  \label{fig_data_rates}
\end{figure}

To investigate the effect of a given threshold choice on the full dataset we ran every event in the DECO database ($\sim$45,000 images) through the CNN model and used the resulting output probabilities to classify each event according to several different threshold choices. The resulting distributions for all event types, shown in Figure~\ref{fig_data_rates}, confirm that a threshold of 0.8 is indeed reasonable and results in ambiguous images $\sim$10\% of the time, which is consistent with human categorization ambiguity (Section~\ref{sec_labels}). With this in mind, the classification scheme based on a threshold of 0.8 was implemented in the public database, which can now be queried by event type as determined by the CNN~\cite{deco_data}. 

\begin{figure}[h]
  \centering
  \includegraphics[width=1\linewidth]{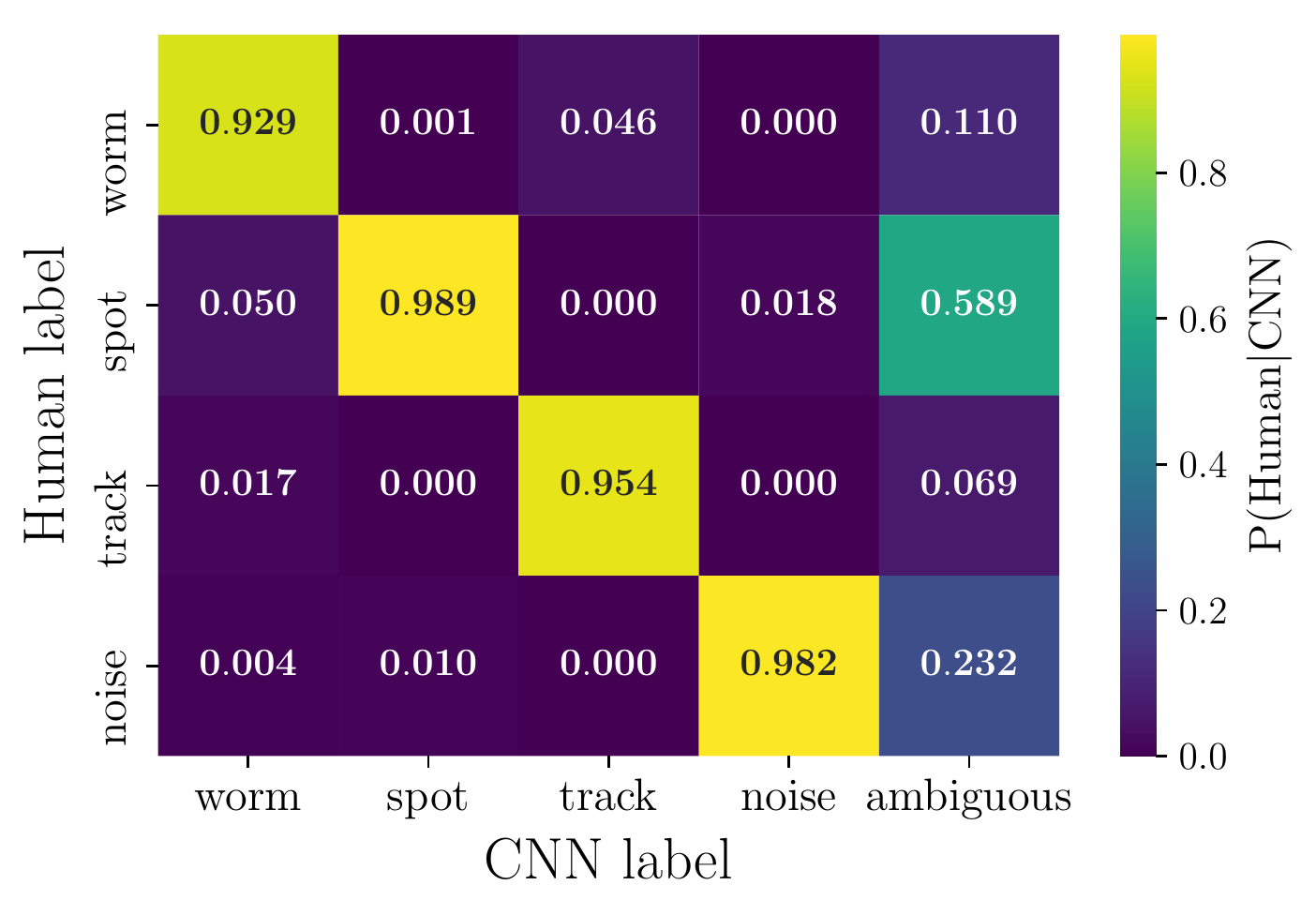}
  \caption{\small{Column-normalized confusion matrix re-weighted to account for the relative rate of each event type in the full data set. In order for an event to be classified as a particular category, the corresponding CNN probability must be $>$ 0.8. Events that do not meet this threshold for any probability are classified as ``ambiguous''.}}
  \label{fig_cm_reweighted}
\end{figure}

Given the classification assigned to any event using this scheme, it is desirable to know the probability that the CNN classification is in fact correct for each event type. As an example, for tracks this corresponds to the conditional probability $P(H=track|CNN=track)$, where $H$ is the human label and $CNN$ is the CNN label. This probability depends on the relative rate of each event type in the data set, i.e., the prior probability that a given event belongs to a given category. The conditional probability could be calculated directly from the testing data sets used in the 10-fold cross validation, however, the distribution of event types in this set of images is biased in comparison to the full dataset. This is because the training set was intentionally enriched with tracks and worms; tracks are the most interesting events from an astrophysical perspective and worms are the primary source of confusion for tracks. Compared to the training set, the full data set has relatively fewer worms and tracks and more spots and noise events. Fortunately, this bias can be corrected by rescaling the testing set results. To accomplish this, we begin with the approximation that the CNN classifications for the full dataset are entirely correct, an approximation that is justified by the excellent performance of the CNN. We then use the abundance of each event type in the full dataset according to the CNN classification to determine the prior probability that an event belongs to a given category. Next, we apply a threshold cut of 0.8 to the testing set and construct a new confusion matrix (similar to Figure~\ref{conf_mat_counts}). We rescale each row of this confusion matrix by the ratio of the number of events for each event type in the full data set (Figure~\ref{fig_data_rates} with a 0.8 threshold) to the number of each event type in the training set (Figure~\ref{samples}). Finally, we rescale the confusion matrix column-wise in order to calculate the conditional probability, $P(H=i|CNN=j)$, for each category. By necessity, a 5th column for “ambiguous” events was added to the confusion matrix, which shows the distribution  of events that don't meet any of the CNN threshold requirements. The resulting confusion matrix, shown in Figure ~\ref{fig_cm_reweighted}, suggests that all four event types in the full dataset are likely to be classified correctly $\geq$ 90\% of the time. Most notably, we estimate that an event classified as a track by the CNN has a $\sim$95\% probability of being a track according to human classification. Note that this quantity is the expected observable purity in the dataset, which differs from the 91\% purity estimated on the training set that was described at the beginning of this section.

\section{Conclusions and Future Work}\label{sec_conclusion}
We have described the development and validation of a convolutional neural network for the classification of images obtained by users running the DECO application. This new approach to image classification resulted in significant improvements over previous classification of DECO images using straight cuts. Event classification using the straight-cuts approach produced a track sample with 20\% purity after applying the rescaling procedure described in Section~\ref{sec_dataset}. The CNN model, on the other hand, yields a data set with an estimated purity of 95\% after rescaling to the full DECO data set. This classification algorithm has been integrated into the standard DECO processing pipeline and the resulting classification of each event is available along with the event's image and metadata on the public web site within several hours of detection. The CNN classification can be used in queries, allowing users to select a sample of images of any particle identity, or multiple identities, for analysis and outreach purposes.

	In addition to improving the overall experience of DECO users, the new model opens the door for new and improved analyses. For example, the model provides efficient rejection of the radioactive background (i.e., worms), which is necessary to detect extensive air showers using DECO or a similar application. Additionally, the measurement of the depletion depth (i.e., sensitive region) of a phone’s camera sensor requires a large, pure sample of cosmic-ray muon tracks. Without a robust method of identifying tracks, the analysis published in ~\cite{vandenbroucke2016} was limited to a single phone. The new classification enables us to extend this analysis to multiple phones with a lower non-cosmic-ray background in the data set. Once the thickness of the depletion region is known for a particular phone model, it can be used to constrain the incident zenith angle of individual cosmic rays. Together with the azimuthal direction of the track within the sensor plane, this will enable reconstructing the direction of DECO tracks. Constraining the direction of detected muons would improve the sensitivity of a multi-phone coincidence analysis, since the direction of muons from the same extensive air shower should be correlated. Measuring the direction of events could also enable measurement of the East-West effect. 
    
    One shortcoming of the analysis presented in this paper is the human labeling method of assembling a sample of training images. There is an inherent bias in the model due to potentially mis-labeled images in the training sample. Although the efforts described in Section~\ref{sec_training} should mitigate some of this bias, further work could quantify it. Beam line data from a particle physics accelerator and data collected from running DECO with radioactive sources would yield unbiased samples of tracks and worms, respectively, to further evaluate the performance of the model. Additionally, coincidence experiments with DECO and scintillators could provide a similar data set of tagged cosmic-ray tracks, though with far lower statistics.
    
    While the model was developed exclusively using images in the Android DECO data set, we expect it to generalize to similar data sets with minimal changes. DECO for iOS, which is currently in development, will have a data set consisting of images created by the same charged-particle interactions discussed here. Although the overall camera response will differ from Android phones, the resulting event types are expected to be the same. It is worth emphasizing that the Android data set consists of images from hundreds of different phone models, with wide variation in camera sensor response to DECO events. The data augmentation applied during training (Section~\ref{sec_data_aug}) mitigates the effects of model-to-model variation by building invariances into the classification that should enable it to generalize to the iOS data set. It is also possible that including the phone model as a feature in the neural network could help further reduce the effects of model-to-model variation. The excellent performance of our CNN in identifying particle types in the DECO data set indicates that the same approach would be powerful for identifying particles detected by other projects that use distributed camera sensors. Finally, our approach (and perhaps our particular model architecture) could be well suited for other experiments (such as the DAMIC~\cite{damic} dark matter project) that use CCD and CMOS sensors for particle detection.

\section*{Acknowledgements}\label{sec_acknowledgments}
DECO is supported by the American Physical Society, the Knight Foundation, the Simon Strauss Foundation, QuarkNet, and by National Science Foundation Grant \#1707945. We are grateful for beta testing, software development, and valuable conversations with Colin Adams, Raaha Azfar, Keith Bechtol, Segev BenZvi, Andy Biewer, Paul Brink, Patricia Burchat, Duncan Carlsmith, Alex Drlica-Wagner, Mike Duvernois, Brett Fisher, Lucy Fortson, Stefan Funk, Mandeep Gill, Laura Gladstone, Giorgio Gratta, Jim Haugen, Kenny Jensen, Kyle Jero, Peter Karn, David Kirkby, Matthew Plewa, David Saltzberg, Marcos Santander, Delia Tosi, and Ian Wisher.
	We would also like to thank Ilhan Bok, Adrian Cisneros, Alex Diebold, Tyler Dolan, Blake Gallay, Emmanuelle Hannibal, Heather Levi, and Owen Roszkowski for their contributions to the DECO project through our QuarkNet DECO high school internship program.

\bibliography{DECO_CNN}

\begin{thebibliography}{10}
\expandafter\ifx\csname url\endcsname\relax
  \def\url#1{\texttt{#1}}\fi
\expandafter\ifx\csname urlprefix\endcsname\relax\def\urlprefix{URL }\fi
\expandafter\ifx\csname href\endcsname\relax
  \def\href#1#2{#2} \def\path#1{#1}\fi

\bibitem{groom2002}
D.~Groom, \href{http://dx.doi.org/10.1023/A:1026196806990}{Cosmic rays and
  other nonsense in astronomical {CCD} imagers}, Experimental Astronomy 14~(1)
  (2002) 45--55.
\newblock \href {http://dx.doi.org/10.1023/A:1026196806990}
  {\path{doi:10.1023/A:1026196806990}}.
\newline\urlprefix\url{http://dx.doi.org/10.1023/A:1026196806990}

\bibitem{vandenbroucke2015}
J.~Vandenbroucke, S.~Bravo, P.~Karn, M.~Meehan, M.~Plewa, T.~Ruggles,
  D.~Schultz, J.~Peacock, A.~L. Simons, {Detecting particles with cell phones:
  the Distributed Electronic Cosmic-ray Observatory}, PoS ICRC2015 (2016) 691.
\newblock \href {http://arxiv.org/abs/1510.07665} {\path{arXiv:1510.07665}}.

\bibitem{crayfis_2016}
D.~Whiteson, M.~Mulhearn, C.~Shimmin, K.~Cranmer, K.~Brodie, D.~Burns,
  \href{http://www.sciencedirect.com/science/article/pii/S0927650516300147}{Searching
  for ultra-high energy cosmic rays with smartphones}, Astroparticle Physics 79
  (2016) 1 -- 9.
\newblock \href
  {http://dx.doi.org/https://doi.org/10.1016/j.astropartphys.2016.02.002}
  {\path{doi:https://doi.org/10.1016/j.astropartphys.2016.02.002}}.
\newline\urlprefix\url{http://www.sciencedirect.com/science/article/pii/S0927650516300147}

\bibitem{unger_2015}
M.~Unger, G.~Farrar, {(In)Feasability of Studying Ultra-High-Energy Cosmic Rays
  with Smartphones} (2015).
\newblock \href {http://arxiv.org/abs/1505.04777} {\path{arXiv:1505.04777}}.

\bibitem{cogliati_2014}
J.~J. Cogliati, K.~W. Derr, J.~Wharton, {Using CMOS Sensors in a Cellphone for
  Gamma Detection and Classification} (2014).
\newblock \href {http://arxiv.org/abs/1401.0766} {\path{arXiv:1401.0766}}.

\bibitem{credo_2017}
P.~Homola, et~al.,
  \href{https://inspirehep.net/record/1667898/files/1804.05614.pdf}{{Search for
  Extensive Photon Cascades with the Cosmic-Ray Extremely Distributed
  Observatory}}, in: {Photon 2017: International Conference on the Structure
  and the Interactions of the Photon and 22th International Workshop on
  Photon-Photon Collisions and the International Workshop on High Energy Photon
  Colliders CERN, Geneva, Switzerland, May 22-26, 2017}, 2018.
\newblock \href {http://arxiv.org/abs/1804.05614} {\path{arXiv:1804.05614}}.
\newline\urlprefix\url{https://inspirehep.net/record/1667898/files/1804.05614.pdf}

\bibitem{vandenbroucke2016}
J.~Vandenbroucke, S.~BenZvi, S.~Bravo, K.~Jensen, P.~Karn, M.~Meehan,
  J.~Peacock, M.~Plewa, T.~Ruggles, M.~Santander, D.~Schultz, A.~Simons,
  D.~Tosi, \href{http://stacks.iop.org/1748-0221/11/i=04/a=P04019}{Measurement
  of cosmic-ray muons with the {Distributed Electronic Cosmic-ray Observatory},
  a network of smartphones}, Journal of Instrumentation 11~(04) (2016) P04019.
\newline\urlprefix\url{http://stacks.iop.org/1748-0221/11/i=04/a=P04019}

\bibitem{meehan2017}
M.~Meehan, S.~Bravo, F.~Campos, J.~Peacock, T.~Ruggles, C.~Schneider, A.~L.
  Simons, J.~Vandenbroucke, M.~Winter, {The particle detector in your pocket:
  The Distributed Electronic Cosmic-ray Observatory}, in: {Proceedings, 35th
  International Cosmic Ray Conference (ICRC 2017): Bexco, Busan, Korea, July
  12-20, 2017}, 2017.
\newblock \href {http://arxiv.org/abs/1708.01281} {\path{arXiv:1708.01281}}.

\bibitem{crayfis_cnn}
M.~Borisyak, M.~Usvyatsov, M.~Mulhearn, C.~Shimmin, A.~Ustyuzhanin, {Muon
  Trigger for Mobile Phones}, J. Phys. Conf. Ser. 898~(3) (2017) 032048.
\newblock \href {http://arxiv.org/abs/1709.08605} {\path{arXiv:1709.08605}},
  \href {http://dx.doi.org/10.1088/1742-6596/898/3/032048}
  {\path{doi:10.1088/1742-6596/898/3/032048}}.

\bibitem{ackermann2012}
M.~Ackermann, et~al., \href{http://stacks.iop.org/0067-0049/203/i=1/a=4}{The
  {Fermi Large Area Telescope} on {Orbit}: {Event Classification, Instrument
  Response Functions}, and {Calibration}}, The Astrophysical Journal Supplement
  Series 203~(1) (2012) 4.
\newline\urlprefix\url{http://stacks.iop.org/0067-0049/203/i=1/a=4}

\bibitem{cms2008}
{The CMS Collaboration},
  \href{http://stacks.iop.org/1748-0221/3/i=08/a=S08004}{The {CMS} experiment
  at the {CERN LHC}}, Journal of Instrumentation 3~(08) (2008) S08004.
\newline\urlprefix\url{http://stacks.iop.org/1748-0221/3/i=08/a=S08004}

\bibitem{deco_data}
\href{https://wipac.wisc.edu/deco}{https://wipac.wisc.edu/deco}.
\newline\urlprefix\url{https://wipac.wisc.edu/deco}

\bibitem{lorensen1987}
W.~E. Lorensen, H.~E. Cline, Marching cubes: A high resolution 3d surface
  construction algorithm, COMPUTER GRAPHICS 21~(4) (1987) 163--169.

\bibitem{scikit-image}
S.~van~der Walt, J.~L. {S}ch\"onberger, J.~{Nunez-Iglesias}, F.~{B}oulogne,
  J.~D. {W}arner, N.~{Y}ager, E.~{G}ouillart, T.~{Y}u, the scikit-image
  contributors, \href{http://dx.doi.org/10.7717/peerj.453}{scikit-image: image
  processing in {P}ython}, PeerJ 2 (2014) e453.
\newblock \href {http://dx.doi.org/10.7717/peerj.453}
  {\path{doi:10.7717/peerj.453}}.
\newline\urlprefix\url{http://dx.doi.org/10.7717/peerj.453}

\bibitem{PDG}
C.~Patrignani, et~al., {Review of Particle Physics}, Chin. Phys. C40~(10)
  (2016) 100001.
\newblock \href {http://dx.doi.org/10.1088/1674-1137/40/10/100001}
  {\path{doi:10.1088/1674-1137/40/10/100001}}.

\bibitem{image_moments}
Y.~D. Khan, S.~A. Khanand, F.~Ahmad, S.~Islam, {Iris Recognition Using Image
  Moments} and k-means {Algorithm}, The Scientific World Journal 2014 (2014) 9.

\bibitem{bengio2009}
Y.~Bengio, \href{http://dx.doi.org/10.1561/2200000006}{{Learning Deep
  Architectures} for {AI}}, Found. Trends Mach. Learn. 2~(1) (2009) 1--127.
\newblock \href {http://dx.doi.org/10.1561/2200000006}
  {\path{doi:10.1561/2200000006}}.
\newline\urlprefix\url{http://dx.doi.org/10.1561/2200000006}

\bibitem{rosenblatt62}
F.~Rosenblatt, Principles of Neurodynamics: Perceptrons and the Theory of Brain
  Mechanisms, Spartan Books, Washington, 1962.

\bibitem{reed1998}
R.~D. Reed, R.~J. Marks, Neural Smithing: Supervised Learning in Feedforward
  Artificial Neural Networks, MIT Press, Cambridge, MA, USA, 1998.

\bibitem{goodfellow2016}
I.~Goodfellow, Y.~Bengio, A.~Courville, Deep Learning, MIT Press, 2016,
  \url{http://www.deeplearningbook.org}.

\bibitem{nair2010}
V.~Nair, G.~E. Hinton,
  \href{http://dl.acm.org/citation.cfm?id=3104322.3104425}{{Rectified Linear
  Units Improve Restricted Boltzmann Machines}}, in: Proceedings of the 27th
  International Conference on International Conference on Machine Learning,
  ICML'10, Omnipress, USA, 2010, pp. 807--814.
\newline\urlprefix\url{http://dl.acm.org/citation.cfm?id=3104322.3104425}

\bibitem{maas2013}
A.~L. Maas, A.~Y. Hannun, A.~Y. Ng, Rectifier nonlinearities improve neural
  network acoustic models, in: in ICML Workshop on Deep Learning for Audio,
  Speech and Language Processing, 2013.

\bibitem{rumelhart1986}
D.~E. Rumelhart, G.~E. Hinton, R.~J. Williams,
  \href{http://dl.acm.org/citation.cfm?id=104279.104293}{Parallel distributed
  processing: Explorations in the microstructure of cognition, vol. 1}, MIT
  Press, Cambridge, MA, USA, 1986, Ch. Learning Internal Representations by
  Error Propagation, pp. 318--362.
\newline\urlprefix\url{http://dl.acm.org/citation.cfm?id=104279.104293}

\bibitem{lecun1998c}
Y.~LeCun, L.~Bottou, G.~B. Orr, K.-R. M\"{u}ller,
  \href{http://dl.acm.org/citation.cfm?id=645754.668382}{Efficient backprop},
  in: Neural Networks: Tricks of the Trade, This Book is an Outgrowth of a 1996
  NIPS Workshop, Springer-Verlag, London, UK, UK, 1998, pp. 9--50.
\newline\urlprefix\url{http://dl.acm.org/citation.cfm?id=645754.668382}

\bibitem{bottou2016}
L.~{Bottou}, F.~E. {Curtis}, J.~{Nocedal}, {Optimization Methods for
  Large-Scale Machine Learning} (Jun. 2016).
\newblock \href {http://arxiv.org/abs/1606.04838} {\path{arXiv:1606.04838}}.

\bibitem{lecun1998a}
Y.~LeCun, L.~Bottou, Y.~Bengio, P.~Haffner, Gradient-based learning applied to
  document recognition, Proceedings of the IEEE 86~(11) (1998) 2278--2323.
\newblock \href {http://dx.doi.org/10.1109/5.726791}
  {\path{doi:10.1109/5.726791}}.

\bibitem{simard2003}
P.~Y. Simard, D.~Steinkraus, J.~C. Platt,
  \href{http://dl.acm.org/citation.cfm?id=938980.939477}{Best practices for
  convolutional neural networks applied to visual document analysis}, in:
  Proceedings of the Seventh International Conference on Document Analysis and
  Recognition - Volume 2, ICDAR '03, IEEE Computer Society, Washington, DC,
  USA, 2003, pp. 958--.
\newline\urlprefix\url{http://dl.acm.org/citation.cfm?id=938980.939477}

\bibitem{boureau2010}
Y.~L. Boureau, J.~Ponce, Y.~Lecun, A {Theoretical Analysis} of {Feature
  Pooling} in {Visual Recognition}, in: ICML 2010 - Proceedings, 27th
  International Conference on Machine Learning, 2010, pp. 111--118.

\bibitem{zhou1988}
Y.~T. Zhou, R.~Chellappa, Computation of optical flow using a neural network,
  in: IEEE 1988 International Conference on Neural Networks, 1988, pp. 71--78
  vol.2.
\newblock \href {http://dx.doi.org/10.1109/ICNN.1988.23914}
  {\path{doi:10.1109/ICNN.1988.23914}}.

\bibitem{willett2013}
K.~W. {Willett}, C.~J. {Lintott}, S.~P. {Bamford}, K.~L. {Masters}, B.~D.
  {Simmons}, K.~R.~V. {Casteels}, E.~M. {Edmondson}, L.~F. {Fortson},
  S.~{Kaviraj}, W.~C. {Keel}, T.~{Melvin}, R.~C. {Nichol}, M.~J. {Raddick},
  K.~{Schawinski}, R.~J. {Simpson}, R.~A. {Skibba}, A.~M. {Smith}, D.~{Thomas},
  {Galaxy Zoo 2: detailed morphological classifications for 304 122 galaxies
  from the Sloan Digital Sky Survey}, MNRAS 435 (2013) 2835--2860.
\newblock \href {http://arxiv.org/abs/1308.3496} {\path{arXiv:1308.3496}},
  \href {http://dx.doi.org/10.1093/mnras/stt1458}
  {\path{doi:10.1093/mnras/stt1458}}.

\bibitem{dieleman2015}
S.~{Dieleman}, K.~W. {Willett}, J.~{Dambre}, {Rotation-invariant convolutional
  neural networks for galaxy morphology prediction}, MNRAS 450 (2015)
  1441--1459.
\newblock \href {http://arxiv.org/abs/1503.07077} {\path{arXiv:1503.07077}},
  \href {http://dx.doi.org/10.1093/mnras/stv632}
  {\path{doi:10.1093/mnras/stv632}}.

\bibitem{ayres2007}
D.~S. Ayres, et~al., {The NOvA Technical Design Report}\href
  {http://dx.doi.org/10.2172/935497} {\path{doi:10.2172/935497}}.

\bibitem{aurisano2016}
A.~{Aurisano}, A.~{Radovic}, D.~{Rocco}, A.~{Himmel}, M.~D. {Messier},
  E.~{Niner}, G.~{Pawloski}, F.~{Psihas}, A.~{Sousa}, P.~{Vahle}, {A
  convolutional neural network neutrino event classifier}, Journal of
  Instrumentation 11 (2016) P09001.
\newblock \href {http://arxiv.org/abs/1604.01444} {\path{arXiv:1604.01444}},
  \href {http://dx.doi.org/10.1088/1748-0221/11/09/P09001}
  {\path{doi:10.1088/1748-0221/11/09/P09001}}.

\bibitem{lecun1998b}
Y.~LeCun, Y.~Bengio,
  \href{http://dl.acm.org/citation.cfm?id=303568.303704}{{The Handbook} of
  {Brain Theory} and {Neural Networks}}, MIT Press, Cambridge, MA, USA, 1998,
  Ch. Convolutional Networks for Images, Speech, and Time Series, pp. 255--258.
\newline\urlprefix\url{http://dl.acm.org/citation.cfm?id=303568.303704}

\bibitem{gong2014}
Y.~{Gong}, L.~{Wang}, R.~{Guo}, S.~{Lazebnik}, {Multi-scale Orderless Pooling
  of Deep Convolutional Activation Features} (Mar. 2014).
\newblock \href {http://arxiv.org/abs/1403.1840} {\path{arXiv:1403.1840}}.

\bibitem{scherer2010}
D.~Scherer, A.~M\"{u}ller, S.~Behnke,
  \href{http://dl.acm.org/citation.cfm?id=1886436.1886447}{Evaluation of
  pooling operations in convolutional architectures for object recognition},
  in: Proceedings of the 20th International Conference on Artificial Neural
  Networks: Part III, ICANN'10, Springer-Verlag, Berlin, Heidelberg, 2010, pp.
  92--101.
\newline\urlprefix\url{http://dl.acm.org/citation.cfm?id=1886436.1886447}

\bibitem{xu2014}
Y.~{Xu}, T.~{Xiao}, J.~{Zhang}, K.~{Yang}, Z.~{Zhang}, {Scale-Invariant
  Convolutional Neural Networks} (Nov. 2014).
\newblock \href {http://arxiv.org/abs/1411.6369} {\path{arXiv:1411.6369}}.

\bibitem{simonyan2014}
K.~{Simonyan}, A.~{Zisserman}, {Very Deep Convolutional Networks for
  Large-Scale Image Recognition} (Sep. 2014).
\newblock \href {http://arxiv.org/abs/1409.1556} {\path{arXiv:1409.1556}}.

\bibitem{krizhevsky2012}
A.~Krizhevsky, I.~Sutskever, G.~E. Hinton,
  \href{http://dl.acm.org/citation.cfm?id=2999134.2999257}{Imagenet
  classification with deep convolutional neural networks}, in: Proceedings of
  the 25th International Conference on Neural Information Processing Systems -
  Volume 1, NIPS'12, Curran Associates Inc., USA, 2012, pp. 1097--1105.
\newline\urlprefix\url{http://dl.acm.org/citation.cfm?id=2999134.2999257}

\bibitem{marcos2016}
D.~{Marcos}, M.~{Volpi}, D.~{Tuia}, {Learning rotation invariant convolutional
  filters for texture classification} (Apr. 2016).
\newblock \href {http://arxiv.org/abs/1604.06720} {\path{arXiv:1604.06720}}.

\bibitem{lenc2014}
K.~{Lenc}, A.~{Vedaldi}, {Understanding image representations by measuring
  their equivariance and equivalence} (Nov. 2014).
\newblock \href {http://arxiv.org/abs/1411.5908} {\path{arXiv:1411.5908}}.

\bibitem{keras}
F.~Chollet, et~al., Keras, \url{https://github.com/fchollet/keras} (2015).

\bibitem{scipy}
E.~Jones, T.~Oliphant, P.~Peterson, et~al.,
  \href{http://www.scipy.org/}{{SciPy}: Open source scientific tools for
  {Python}} (2001--).
\newline\urlprefix\url{http://www.scipy.org/}

\bibitem{kukacka2017}
J.~{Kuka{\v c}ka}, V.~{Golkov}, D.~{Cremers}, {Regularization for Deep
  Learning: A Taxonomy} (Oct. 2017).
\newblock \href {http://arxiv.org/abs/1710.10686} {\path{arXiv:1710.10686}}.

\bibitem{szegedy2015}
C.~{Szegedy}, V.~{Vanhoucke}, S.~{Ioffe}, J.~{Shlens}, Z.~{Wojna}, {Rethinking
  the Inception Architecture for Computer Vision} (Dec. 2015).
\newblock \href {http://arxiv.org/abs/1512.00567} {\path{arXiv:1512.00567}}.

\bibitem{hinton2012}
G.~E. {Hinton}, N.~{Srivastava}, A.~{Krizhevsky}, I.~{Sutskever}, R.~R.
  {Salakhutdinov}, {Improving neural networks by preventing co-adaptation of
  feature detectors} (Jul. 2012).
\newblock \href {http://arxiv.org/abs/1207.0580} {\path{arXiv:1207.0580}}.

\bibitem{srivastava2014}
N.~Srivastava, G.~Hinton, A.~Krizhevsky, I.~Sutskever, R.~Salakhutdinov,
  \href{http://jmlr.org/papers/v15/srivastava14a.html}{Dropout: A simple way to
  prevent neural networks from overfitting}, Journal of Machine Learning
  Research 15 (2014) 1929--1958.
\newline\urlprefix\url{http://jmlr.org/papers/v15/srivastava14a.html}

\bibitem{srebro2005}
N.~Srebro, A.~Shraibman, \href{http://dx.doi.org/10.1007/11503415_37}{Rank,
  {Trace-norm} and {Max-norm}}, in: Proceedings of the 18th Annual Conference
  on Learning Theory, COLT'05, Springer-Verlag, Berlin, Heidelberg, 2005, pp.
  545--560.
\newblock \href {http://dx.doi.org/10.1007/11503415\_37}
  {\path{doi:10.1007/11503415\_37}}.
\newline\urlprefix\url{http://dx.doi.org/10.1007/11503415_37}

\bibitem{bishop1995}
C.~M. {Bishop}, Regularization and {Complexity Control} in {Feed-forward
  Networks}, in: F.~Fougelman-Soulie, P.~Gallinari (Eds.), Proceedings
  International Conference on Artificial Neural Networks ICANN'95, Vol.~1,
  1995, pp. 141--148.

\bibitem{sjoberg1995}
J.~{Sj{\"o}berg}, L.~{Ljung},
  \href{http://www.sciencedirect.com/science/article/pii/S1474667017507156}{Overtraining,
  {Regularization}, and {Searching} for {Minimum} in {Neural Networks}}, IFAC
  Proceedings Volumes 25~(14) (1992) 73 -- 78, 4th IFAC Symposium on Adaptive
  Systems in Control and Signal Processing 1992, Grenoble, France, 1-3 July.
\newblock \href
  {http://dx.doi.org/https://doi.org/10.1016/S1474-6670(17)50715-6}
  {\path{doi:https://doi.org/10.1016/S1474-6670(17)50715-6}}.
\newline\urlprefix\url{http://www.sciencedirect.com/science/article/pii/S1474667017507156}

\bibitem{zeiler2012}
M.~D. {Zeiler}, {ADADELTA: An Adaptive Learning Rate Method} (Dec. 2012).
\newblock \href {http://arxiv.org/abs/1212.5701} {\path{arXiv:1212.5701}}.

\bibitem{kingma2014}
D.~P. {Kingma}, J.~{Ba}, {Adam: A Method for Stochastic Optimization} (Dec.
  2014).
\newblock \href {http://arxiv.org/abs/1412.6980} {\path{arXiv:1412.6980}}.

\bibitem{theano}
{Theano Development Team}, \href{http://arxiv.org/abs/1605.02688}{{Theano: A
  {Python} framework for fast computation of mathematical expressions}}
  arXiv:1605.02688.
\newline\urlprefix\url{http://arxiv.org/abs/1605.02688}

\bibitem{kohavi1995}
R.~Kohavi, A study of cross-validation and bootstrap for accuracy estimation
  and model selection, Morgan Kaufmann, 1995, pp. 1137--1143.

\bibitem{damic}
A.~E. Chavarria, et~al., Damic at snolab, Physics Procedia 61 (2015) 21 -- 33,
  13th International Conference on Topics in Astroparticle and Underground
  Physics, TAUP 2013.
\newblock \href {http://dx.doi.org/https://doi.org/10.1016/j.phpro.2014.12.006}
  {\path{doi:https://doi.org/10.1016/j.phpro.2014.12.006}}.

\end{thebibliography}
\end{document}